\documentclass[twocolumn,aps,prd,showpacs,superscriptaddress,amsmath,amssymb,nofootinbib,nobibnotes]{revtex4}

\usepackage{anyfontsize}
\usepackage{amssymb}
\usepackage{amsmath}
\usepackage{amsfonts}
\usepackage{amsbsy}
\usepackage[utf8]{inputenc} 
\usepackage{newtxtext}
\usepackage{newtxmath}
\usepackage{gensymb}

\usepackage{tensor}
\usepackage{mathrsfs}

\usepackage{bm}

\usepackage[utf8]{inputenc}

\usepackage{graphicx}
\usepackage{epsfig}
\usepackage{epstopdf}

\usepackage[usenames,dvipsnames]{xcolor}

\usepackage{verbatim,mathtools,needspace,enumitem,etoolbox}

\definecolor{linkcolor}{rgb}{0.0,0.3,0.5}

\usepackage[unicode, colorlinks=true, linkcolor=linkcolor, citecolor=linkcolor, filecolor=linkcolor,urlcolor=linkcolor, pdfusetitle]{hyperref}

\usepackage{epstopdf}

\usepackage{bm}
\usepackage{dcolumn}

\usepackage{longtable}
\usepackage{enumerate}
 \usepackage[normalem]{ulem}

\hypersetup{colorlinks=true,
            citecolor=blue,
            linkcolor=blue,
            urlcolor=blue}
\usepackage{afterpage}
\usepackage[T1]{fontenc}
\usepackage{subfigure}
\usepackage{amsmath}
\usepackage{float}
\usepackage{mathrsfs}
\usepackage[dvipsnames]{xcolor}
\usepackage{soul}

\definecolor{darkgreen}{RGB}{0,201,0}

\setcitestyle{numbers,square}

\setcitestyle{numbers,square}

\begin{document}
\title{Shadows and lensing of black holes immersed in strong magnetic fields}

\author{Haroldo C. D. Lima Junior}
\email{haroldolima@ufpa.br} 
\affiliation{Faculdade de F\'{\i}sica,
Universidade Federal do Par\'a, 66075-110, Bel\'em, PA, Brazil }

\author{Pedro V. P. Cunha}%
 \email{pvcunha@ua.pt}
\affiliation{Departamento de Matem\'atica da Universidade de Aveiro and Centre for Research and Development  in Mathematics and Applications (CIDMA), Campus de Santiago, 3810-183 Aveiro, Portugal }

\author{Carlos A. R. Herdeiro}
\email{herdeiro@ua.pt} 
\affiliation{Departamento de Matem\'atica da Universidade de Aveiro and Centre for Research and Development  in Mathematics and Applications (CIDMA), Campus de Santiago, 3810-183 Aveiro, Portugal }

\author{Lu\'{\i}s C. B. Crispino}
\email{crispino@ufpa.br} 
\affiliation{Faculdade de F\'{\i}sica,
Universidade Federal do Par\'a, 66075-110, Bel\'em, PA, Brazil }

\date{\today}

\begin{abstract}
%
 We investigate the null geodesic flow and in particular the light rings (LRs), fundamental photon orbits (FPOs) and shadows of a black hole (BH) immersed in a strong, uniform magnetic field, described by the Schwarzschilld-Melvin electrovacuum solution. The empty Melvin magnetic Universe contains a tube of planar  LRs. Including a BH, for weak magnetic fields, the shadow becomes oblate, whereas the intrinsic horizon geometry becomes prolate. For strong magnetic fields (overcritical solutions), there are no LRs outside the BH horizon, a result explained using topological arguments. This feature, together with the light confining structure of the Melvin universe yields  \textit{panoramic shadows}, seen (almost) all around the equator of the observer's sky. Despite the lack of LRs, there are FPOs,  including polar planar ones, which define the shadow  edge.  We also observe and discuss chaotic lensing, including in the empty Melvin universe, and multiple disconnected shadows.  
\end{abstract}

\maketitle

\section{Introduction}
The bending of light rays due to a gravitational field is one of the most important predictions of general relativity (GR). For a body with the Sun's mass, the deflection, first predicted by Einstein, is less than two arcseconds. After some frustrated attempts~\cite{LCBC::2020}, the experimental measurement of the bending of a light ray passing close to the Sun was first accomplished during the 1919 total solar eclipse. The deviation angle was measured to be approximately 1.75 arcseconds, and it was in agreement with the values predicted by GR~\cite{Dyson} (cf. Ref.~\cite{LCBC::2019} for a historical account).  
Higher deflection angles (around ten arcseconds) are obtained for the light emitted by quasars (active galactic nuclei with high luminosity, discovered during the 1960s) and deflected by galaxies along the line of sight.

The deflection angles for the Sun and quasars are relatively small. For black holes (BHs), these angles can be arbitrarily large, and light rays can even describe planar circular closed orbits dubbed in the literature as light rings (LRs). It has been proven that stationary, axisymmetric and asymptotically flat BHs must have one ``standard" LR outside the event horizon for each rotation sense~\cite{Cunha::2020}.\footnote{See also~\cite{Cunha::2017,Wei::2020,Guo:2020qwk}.}
Such LRs are directly related to the shadows of BHs~\cite{Cunha&Herdeiro::2018}. The shadow of a BH~\cite{Falcke:1999pj} is the dark region formed when a BH is illuminated by some light source, for instance by a celestial sphere, as seen by an observer inside this celestial sphere. 
The first proper analysis of the shadow cast by an isolated spinning BH (described by the Kerr metric~\cite{Kerr:1963ud}) was performed by Bardeen~\cite{Kerr_shadow}. Over the last few years, in particular motivated by the Event Horizon Telescope (EHT) collaboration imaging of the shadow of M87*~\cite{EHT}, the computation of BH shadows has become an active area of research, as the means to test GR and modified gravity, the BH paradigm, and to also gain insight into strong gravity features and into new physics, see, $e.g.$, for some recent work, Refs.~\cite{Kerr_Newman_Shadow,Rotating_regularBH_Shadow, KerrwSH_Shadow, Cunha:2019ikd,Cunha:2016bjh, Kerr_Sen_Shadow, Cunha:2019dwb, Cunha:2018uzc, Cunha:2018cof, Cunha:2016wzk,Abdujabbarov:2016hnw,Wei:2020ght,Vincent:2015xta,Bacchini:2021fig, Eichhorn:2021iwq,
Afrin:2021imp,
Glampedakis:2021oie,
Zhang:2021pvx,
Junior:2021atr,
Qian:2021qow,Herdeiro:2021lwl,Lee:2021sws,Volkel:2020xlc,Zhang:2020xub,Contreras:2020kgy,
Ma:2020dhv,Long:2020wqj,Belhaj:2020mlv,Dokuchaev:2020wqk,Khodadi:2020jij,Badia:2020pnh,Ovgun:2020gjz,
Chen:2020qyp,Creci:2020mfg}.

In astrophysically realistic environments, BHs are not isolated. Instead, they are surrounded by matter and fields that can interact with the central BH, giving rise to interesting phenomena. For instance, the magnetic field generated due to matter accretion by the BH allows energy extraction through the so-called Blandford-Znajek mechanism~\cite{Blandford::1977}. Moreover, BHs may be subjected to an external magnetic field coming from a neutron star companion close to the BH. This has been observed to be the case of Sagittarius A* (Sgr A*), in the center of the Milky Way, which is surrounded by a magnetar known as SGR J1745-29~\cite{Kaya Morietal::2013, Kenneaetal::2013, Eatoughetal::2013}. The magnetic field of SGR J1745-29 is $B=1.6\,\times 10^{14}$ gauss, being among the strongest magnetic fields observed so far~\cite{Olausen::2014}. A recently released EHT analysis, moreover, corroborates the existence of strong magnetic fields near the event horizon  of M87*~\cite{EHT2021}. 

These environments, where an external strong magnetic field surrounds a BH, can be modeled analytically with the Schwarzschild-Melvin BH (SMBH) exact solution of Einstein-Maxwell theory, obtained by Ernst~\cite{Ernst::1976}. 
The SMBH solution represents a Schwarzschild BH spacetime immersed in a Melvin universe~\cite{Melvin::1964}. The  latter is a self gravitating ``constant"  magnetic field and it is not asymptotically flat, since the magnetic field extends all the way to spatial infinity. Despite this unphysical feature,  in  what concerns astrophysical BHs, the SMBH geometry can,  nonetheless, be relevant in describing the near horizon geometry of a BH subjected to certain strong magnetic fields.

The SMBH spacetime presents interesting properties, which have been extensively studied. For instance, the motion of charged particles on the equatorial plane of the SMBH spacetime was studied in Refs.~\cite{Dadhich::1979,Karas::1992,Vokrouhlicky::1991,Lim::2015}, while the motion of timelike and null geodesics on the equatorial plane was analyzed in Refs.~\cite{Esteban::1984,Dhurandhar::1983,Stuchlik::1999}, and it was noticed that the SMBH spacetime can have \textit{no LR at all} outside the event horizon. One of the goals of this paper is to explore this intriguing result from the viewpoint of the topological arguments provided in~\cite{Cunha::2020}, that guarantee that asymptotically flat BHs \textit{must} have LRs. Another goal is to understand how this result impacts on the shadow of these BHs.

Rotating Kerr BHs in the presence of a magnetic field were studied by Wald~\cite{Wald::1974}, where the rotation axis is aligned to the magnetic field. Interestingly, this rotating Kerr BH surrounded by an external magnetic field acquires a net electric charge $Q=2BJ$, where $B$ is the strength of the magnetic field and $J$ is the angular momentum of the BH.  Moreover the absorption and scattering of planar scalar waves in the SMBH in the limit of weak magnetic field~\cite{Chen::2013}, the quasinormal modes of the static and rotating BHs in external magnetic fields~\cite{Konoplya::2008, Brito::2014,Konoplya::2008SPI} as well as stationary scalar clouds~\cite{Santos:2021nbf} were also studied.

As the mass parameter of the SMBH goes to zero, the spacetime reduces to the Melvin magnetic Universe~\cite{Melvin::1964}, which was originally proposed as an example of pure electromagnetic geon, i.e. an electromagnetic field held together by its own gravitational field~\cite{Wheeler::1955}. The properties of the motion of charged and uncharged massive particles, as well as photons, in the Melvin spacetime were also investigated in~\cite{Melvin::1966,Thorne::1965,Lim::2020}, which was generalized to Einstein-nonlinear electrodynamics in~\cite{Gibbons:2001sx}.

In this paper we shall analyze the LRs and shadows of static chargeless BHs surrounded by a strong external magnetic field, modeled by the SMBH spacetime. As a particular (vanishing mass  parameter) case, we study the LRs and the gravitational lensing of the Melvin magnetic Universe.
The remainder of this paper is organized as follows. 
In Sec.~\ref{spacetime} we review the properties of the SMBH spacetime. In Sec.~\ref{Null_geodesics} we define a 2D effective potential, study the motion of null geodesics and the LRs in the SMBH spacetime, as well as in the Melvin solution. In Sec.~\ref{Shadows} we study the shadows cast by the SMBH, the gravitational lensing of the Melvin spacetime and present a selection of our backwards ray-tracing results. Our final remarks are presented in Sec.~\ref{Final_remarks}.

\section{The spacetime}
\label{spacetime}
The geometry of the SMBH spacetime, {which is a static and axially symmetric  BH solution of the Einstein-Maxwell equations, is described by}
the following line element \cite{Ernst::1976}:
\begin{align}
\label{line_el}\nonumber ds^2=-&\Lambda^2(r,\theta)\left(1-\frac{2M}{r}\right)\,dt^2+\frac{\Lambda^2(r,\theta)}{\left(1-\frac{2M}{r}\right)}\,dr^2\\+&r^2\Lambda^2(r,\theta)\,d\theta^2+
\frac{r^2\,\sin^2\theta}{\Lambda^2(r,\theta)}\,d\phi^2,
\end{align}
where
\begin{align}
\Lambda(r,\theta)=1+\frac{B^2r^2\sin^2\theta}{4},
\end{align}
$M$ is the mass of the BH and $B$ determines the strength of the external magnetic field.\footnote{In this paper we assume that $B\geqslant 0$. This choice does not imply in any loss of generality.} The electromagnetic potential 1-form is given by~\cite{Lim::2015}
\begin{equation}
A=\frac{Br^2\sin^2\theta}{2\Lambda}d\phi.
\end{equation}
{Due to the external ``constant" magnetic field, the SMBH geometry with $B\neq 0$ does not approach the Minkowski spacetime in the asymptotic region.} For $B=0$, the SMBH spacetime reduces to the (vacuum) Schwarzschild solution, while for $M=0$ it reduces to the Melvin magnetic Universe~\cite{Melvin::1964}. 

{The line element \eqref{line_el} exhibits singularities at $r=0$ and $r=2M$. 
The former corresponds to an irremovable singularity, since the Kretschmann scalar
\begin{equation}
\label{Kretschmann_scalar}K=\frac{\mathcal{P}(r,\theta)}{r^6\Lambda^8},
\end{equation}
diverges at $r=0$, regardless of the value of the angular coordinate $\theta$. In Eq.~\eqref{Kretschmann_scalar}, $\mathcal{P}(r,\theta)$ is a long and nonenlightening function of the coordinates $(r,\theta)$, obeying
\begin{equation}
\lim_{r\rightarrow 0}\mathcal{P}(r,\theta)=48M^2.
\end{equation}
The singularity at $r=2M$ is merely a coordinate one. This can be confirmed by writing the SMBH line element~\eqref{line_el} in terms of the Eddington-Finkelstein-like coordinate $u$, defined through
\begin{equation}
du=dt+\frac{dr}{1-\frac{2M}{r}},
\end{equation}
obtaining that
\begin{align}
\nonumber ds^2=&-\Lambda^2\left(1-\frac{2M}{r}\right)du^2+2\Lambda^2dr\,du+r^2\Lambda^2d\theta^2\\
&+\frac{r^2\,\sin^2\theta}{\Lambda^2}\,d\phi^2,
\end{align}
which is regular at $r=2M$. In the Schwarzschild geometry, $r=2M$ corresponds to the location of the BH event horizon.
The event horizon for asymptotically flat geometries is a well-defined teleological concept: the boundary of  the causal past of future null infinity. Identifying it requires knowledge of the global structure of the spacetime~\cite{HawkingandEllis_book::1973}. However, the formulation for nonasymptotically flat spacetimes may introduce conceptual and technical difficulties. The existence of a BH region in the SMBH spacetime may,  nonetheless, be justified by the identification  of an {\it apparent horizon}, which requires only local knowledge~\cite{HawkingandEllis_book::1973}. To  show $r=2M$ delimits a BH region, in  this sense,  let us denote the tangent vectors of a congruence of outgoing and ingoing null geodesics by $l^\mu$ and $n^\mu$, respectively. These geodesic congruences are not necessarily affinely parametrized, i.e.,
\begin{align}
\label{geoeq_out}&l^\mu\nabla_\mu l^\nu=k_l l^\nu,\\
\label{geoeq_in}&n^\mu\nabla_\mu n^\nu=k_n n^\nu,
\end{align} 
where $k_l,k_n$ are some functions. The expansion $\Theta$ of these null congruences are given by
\begin{align}
\label{expansion_out}&\Theta_{(l)}=\nabla_\mu l^\mu-k_l,\\
\label{expansion_in}&\Theta_{(n)}=\nabla_\mu n^\mu-k_n.
\end{align}
The apparent horizon is defined as the outer boundary of a trapped region, where the following condition holds~\cite{HawkingandEllis_book::1973,Wald,Poisson_book::2004}:
\begin{align}
\label{apparent_horizon1} \Theta_{(l)}=0.
\end{align}
For radially outgoing and ingoing null geodesic congruences along the equatorial plane, we have
\begin{align}
\label{out_congruence}&l^\mu=\left(l^u,l^r,l^\theta,l^\phi\right)=\left(2, 1-\frac{2M}{r},0,0\right),\\
\label{in_congruence}&n^\mu=\left(n^u,n^r,n^\theta,n^\phi\right)=\left(0,-2,0,0\right), 
\end{align}
respectively. Using Eqs.~\eqref{out_congruence} and \eqref{in_congruence} together with Eqs.~\eqref{geoeq_out} and \eqref{geoeq_in}, we find that
\begin{align}
\label{k_l}&k_l=\frac{2M}{r^2}+\frac{B^2\sin\theta^2}{\Lambda}\left(r-2M\right),\\
&k_n=-\frac{2Br\sin\theta^2}{\Lambda}.
\end{align}
Moreover, from Eqs.~\eqref{expansion_out},  \eqref{out_congruence}, and \eqref{k_l}, we have
\begin{align}
&\Theta_{(l)}=\frac{2\left(r-2M\right)}{r^2},
\label{expansion_out2}
\end{align}
which vanishes at $r=2M$, as required for an apparent horizon. For the sake of completeness, we can also show that the expansion of null ingoing geodesics is negative 
\begin{equation}
\label{apparent_horizon2}\Theta_{(n)}=-\frac{4}{r}<0.
\end{equation}
Therefore, we conclude that $r_h=2M$ is the radial coordinate of an apparent horizon in the SMBH spacetime, since at this point  the expansion~\eqref{expansion_out2} is zero and for $r<2M$ it is negative. }

\begin{figure}
\centering
\includegraphics[scale=0.68]{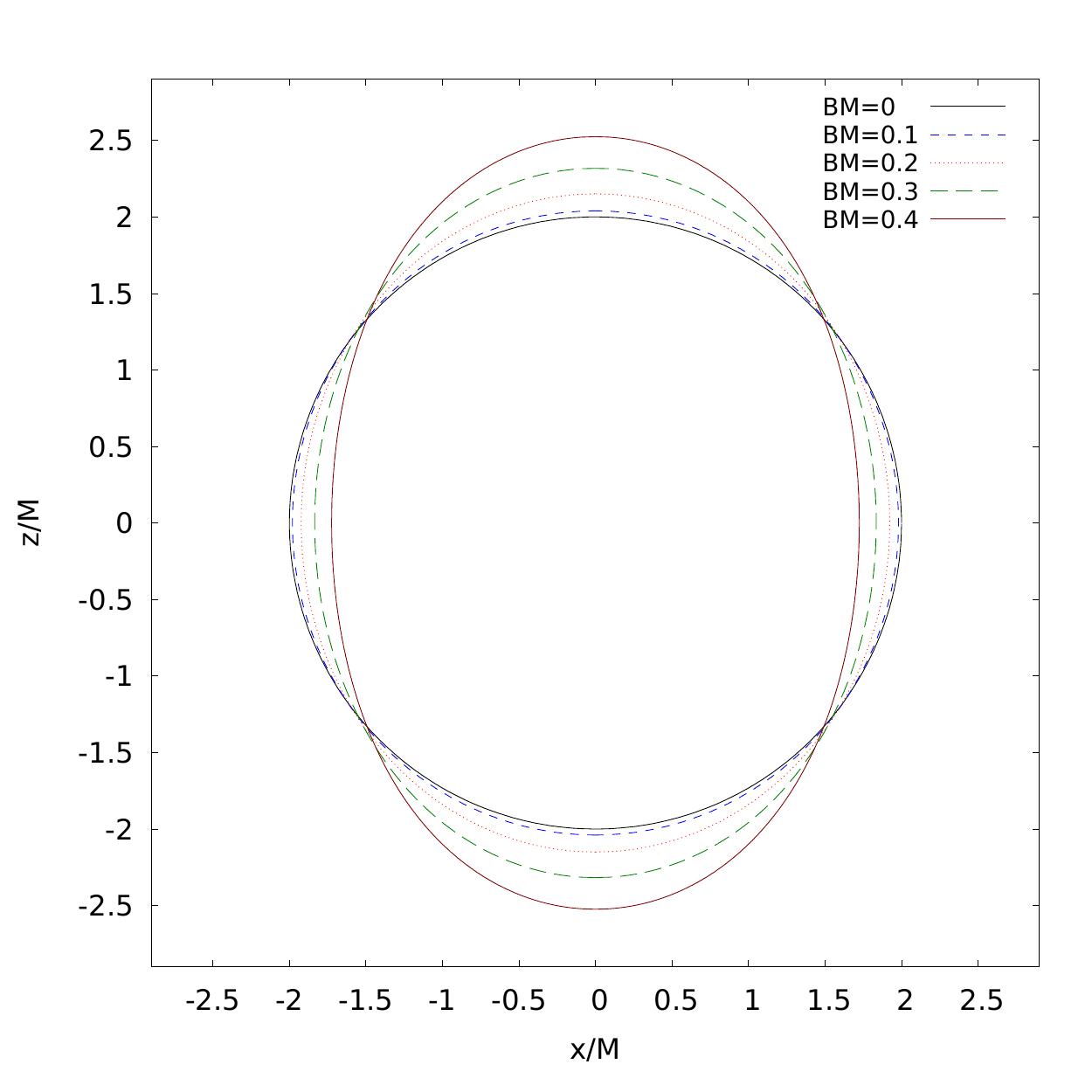}
\caption{Embedding of the {apparent} horizon geometry in Euclidean 3-space for the SMBH spacetime with several values of $B$. For $B=0$, it reduces to the vacuum Schwarzschild case. For $B>0$, the {apparent} horizon geometry becomes  increasingly prolate.} %
\label{Horizon_embedding}
\end{figure}

The surface area of the {apparent} horizon is 
\begin{equation}
A_h=\left.\int_0^{2\pi}\int_0^\pi\sqrt{g_{\theta\theta}g_{\phi\phi}}d\theta d\phi\right|_{r=r_h}=4\pi r_h^2,
\end{equation} 
which is equal to that of the vacuum Schwarzschild geometry. Although the radial coordinate and the surface area of the {apparent} horizon are equal in the SMBH and vacuum Schwarzschild case, the geometry of the {apparent} horizon is different, since the SMBH spacetime is not spherically symmetric. To illustrate this point, in Fig.~\ref{Horizon_embedding} we show the isometric embedding in Euclidean 3-space of the SMBH {apparent} horizon geometry~\cite{Wild::1980,Kulkarni::1986, Gibbons::2009}. The {apparent} horizon keeps a $\mathbb{Z}_ 2$ north-south symmetry with a well-defined equator, but becomes increasingly prolate and flattened near the equator, as $BM$ is increased.  In the next sections, we shall explore the properties of null geodesics on and outside the equatorial plane.

\section{Null geodesics}
\label{Null_geodesics}
\subsection{{LRs in the SMBH (and Melvin) spacetimes}}
\label{sec3a}

The motion of null geodesics can be described by Hamilton's equations:
\begin{align}
\label{Ham1}&\dot{x}^\mu=\frac{\partial \mathcal{H}}{\partial p_\mu},\\
\label{Ham2}&\dot{p}_\mu=-\frac{\partial \mathcal{H}}{\partial x^\mu},
\end{align}
where the Hamiltonian is given by
\begin{equation}
\label{Hamiltonian}\mathcal{H}=\frac{1}{2}g^{\mu\nu}p_\mu p_\nu=0.
\end{equation}
Since the metric components of the SMBH do not depend on the $t$ and $\phi$ coordinates, we have two conserved quantities, namely
\begin{align}
&p_t=-E,\\
&p_\phi=L.
\end{align}
The quantities $\{E,L\}$ are usually interpreted as the energy and angular momentum of the light ray. However, due to nontrivial Melvin asymptotics, we remark that $E$ is not generically equal to the energy measured by an observer at spatial infinity (nor should such an observer  be regarded as preferential). 
For instance, the energy measured by a static observer at a radius $r$ and colatitude $\theta$ is given by
\begin{equation}
\label{Eobs}E_{obs}=-p_\mu {u}^\mu=\frac{E}{\left(1+\frac{B^2r^2\sin^2\theta}{4} \right)\left(1-\frac{2M}{r}\right)^\frac{1}{2}},
\end{equation}
where ${u}^\mu$ is the 4-velocity of the observer. When $B\neq 0$ and the observer lies outside the symmetry axis, $\theta\neq 0,\pi$,  $E_{obs}\rightarrow 0$  as $r\rightarrow \infty$. If the observer lies along the axis of symmetry, however, $E_{obs}=E$ at spatial infinity. Hence, one may  describe $E$  as the energy of the photon as measured by an observer lying on the axis of symmetry, i.e., along the direction of the magnetic field lines, at $r\rightarrow \infty$.

Since the SMBH spacetime is not (generically) spherically symmetric, the general motion cannot be reduced to the %
 equatorial plane. Moreover the Hamilton-Jacobi equation for the SMBH is not known to be separable; hence one is not able (in general) to write the equations of motion as four first order differential equations, as, for instance, in the Kerr case~\cite{Carter:: 1968}. The case $M=0$ (Melvin solution), however, admits separability due to an extra Killing vector related to the translation symmetry along the $z$ direction~\cite{Melvin::1966}.
 
The Hamiltonian $\mathcal{H}$ can be rewritten as a sum of a kinetic term and a potential term, namely:
\begin{align}
\mathcal{H}=T(r,\theta)+V(r,\theta,E,L),
\end{align}
where
\begin{align}
&T(r,\theta)=g^{rr}(p_r)^2+g^{\theta\theta}(p_\theta)^2,\\
&V(r,\theta,E,L)=-\frac{E^2}{\Lambda^2\left(1-\frac{2M}{r}\right)}+\frac{\Lambda^2L^2}{r^2\sin^2\theta}.
\end{align}
Since the potential $V(r,\theta,E,L)$ depends on $E$ and $L$, we define a new \textit{effective potential}, $H(r,\theta)$, which is independent of $E$ and $L$, given by
\begin{equation}
\label{eff_pot}
H(r,\theta)\equiv \frac{\Lambda^2\left(1-\frac{2M}{r}\right)^\frac{1}{2}}{r\sin\theta},
\end{equation}
so that we can write
\begin{equation}
\label{Vtoh}
V(r,\theta,E,L)=\frac{L^2}{\Lambda^2\left(1-\frac{2M}{r}\right)}\left(H(r,\theta)+\frac{1}{\eta}\right)\left(H(r,\theta)-\frac{1}{\eta}\right),\\
\end{equation}
where $\eta\equiv L/E$. 
Since $T\geqslant 0$ and $V\leqslant 0$, the null geodesics in the SMBH spacetime obey the following inequality: 
\begin{equation}
\label{geo_cond}\frac{1}{|\eta|}\geqslant H(r,\theta), 
\end{equation}
where the equality holds at \textit{turning points}.  When $B\neq 0$, as $r\rightarrow \infty$, the effective potential diverges (away from the symmetry axis), $H(r,\theta)\rightarrow \infty$. Thus,  Eq.~\eqref{geo_cond} implies that a generic light ray  (not necessarily restricted to the equatorial plane) can only escape to infinity if $\eta=0$. This property holds for the whole SMBH family, with $B\neq 0$, including the $M=0$ Melvin universe.

Let us now turn our attention to LRs, i.e., planar circular photon orbits, by analyzing the properties of the effective potential \eqref{eff_pot}. LRs are defined by $p_r=p_\theta=0$ and $\dot{p}_r=\dot{p}_\theta=0$.\footnote{As we shall see later, in the SMBH spacetime there are planar null (but geometrically noncircular) orbits which are polar, rather than equatorial - Fig.~\ref{FPO} -, with $p_\theta\neq 0$ e $\eta=0$. These are not LRs and they do not appear as  critical points of $H(r,\theta)$. The tangent vector to these arbits is not a combination of the Killing vector fields $\partial_t$ and $\partial_\phi$, which is the case for LRs~\cite{Cunha::2017}.} They correspond to the critical points of $V$, {\it i.e.}:
\begin{align}
&V=0,\\
&\nabla V=0,
\end{align}
{which may be rewritten in terms of $H(r,\theta)$ as
\begin{align}
\label{LR1}&H(r,\theta)=\frac{1}{\eta},\\
\label{LR2}&\nabla H(r,\theta)=0.
\end{align}
}

\begin{figure}
\centering
\subfigure{\includegraphics[scale=0.68]{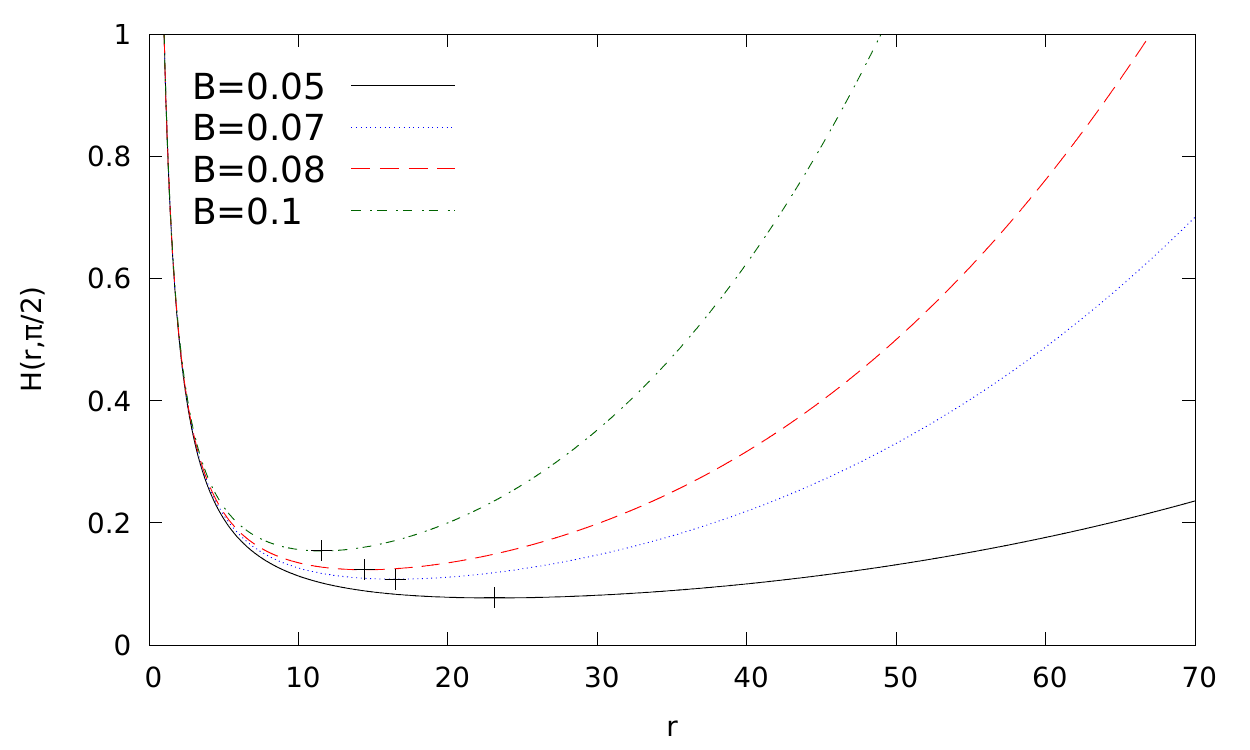}}
\subfigure{\includegraphics[scale=0.68]{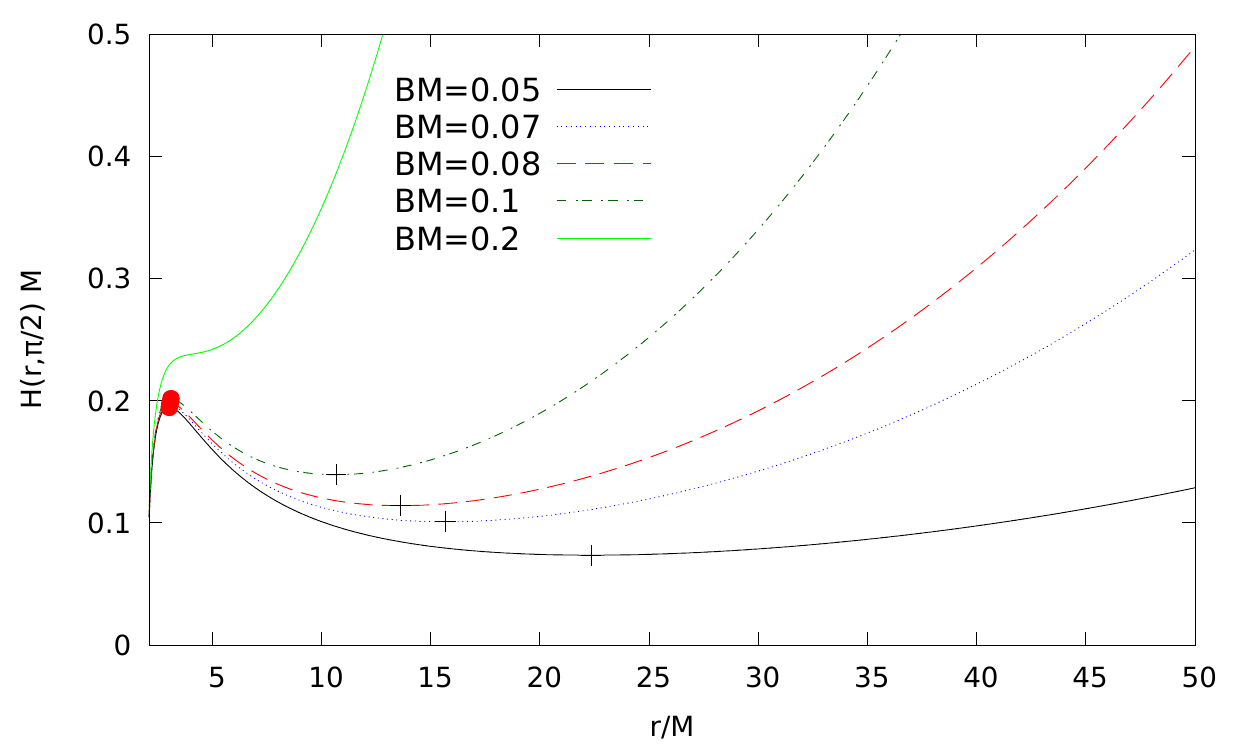}}
\caption{Top: Effective potential {$H(r,\theta)$} for the Melvin spacetime, evaluated at the equatorial plane. Bottom: Effective potential for the SMBH spacetime, also evaluated at the equatorial plane. We have chosen different values of magnetic field strength. The red (black) dots (crosses) correspond to the equatorial unstable (stable) LRs radial coordinate. For $B>B_c$ there are no LRs (green curve). %
}
\label{Veff_equatorial}
\end{figure}

{Consider first the Melvin spacetime, by choosing $M=0$ in Eq.~\eqref{eff_pot}. From Eq.~\eqref{LR2} it is straightforward to identify one equatorial ($\theta=\pi/2$) LR at
\begin{equation}
\label{LR_Melvin}r=\frac{2}{\sqrt{3}B}. 
\end{equation}
This LR, however, is not unique due to the existence of an additional spacelike Killing vector in the Melvin universe.
This can be best seen by writing the Melvin solution in ($t$, $\rho$, $z$, $\phi$) coordinates, where 
\begin{align}
&\rho=r\sin\theta,\\
&z=r\cos\theta.
\end{align}
The line element~\eqref{line_el} becomes
\begin{align}
\label{line_el2}&ds^2=\Lambda^2(\rho)\left(\,-dt^2+d\rho^2+dz^2\right)+\frac{\rho^2}{\Lambda^2(\rho)}d\phi^2,\\
&\Lambda(\rho)=1+\frac{B^2\rho^2}{4},
\end{align}
which is independent of $t$, $z$ and $\phi$. 
The LR \eqref{LR_Melvin}, in the ($t, \rho,z, \phi$) coordinates, is determined by
\begin{align}
\label{rho_LR}&\rho=\frac{2}{\sqrt{3}B},\\
&z=0.
\end{align}
Since the Melvin universe has a translation symmetry along the $z$ direction, there is a LR at each $z=$ constant plane with a cylindrical radius given by $\eqref{rho_LR}$. Hence, we conclude that the Melvin universe admits a \textit{tube of planar LRs}. This means that, instead of just a single LR orbit, there exists an  infinite continuous set of LRs.}
In Fig.~\ref{Veff_equatorial}~(top panel) we show the effective potential {$H(r,\pi/2)$} for the Melvin spacetime.  The \textit{tube of planar LRs} is stable against radial perturbations.

\begin{figure}
\centering
\includegraphics[scale=0.85]{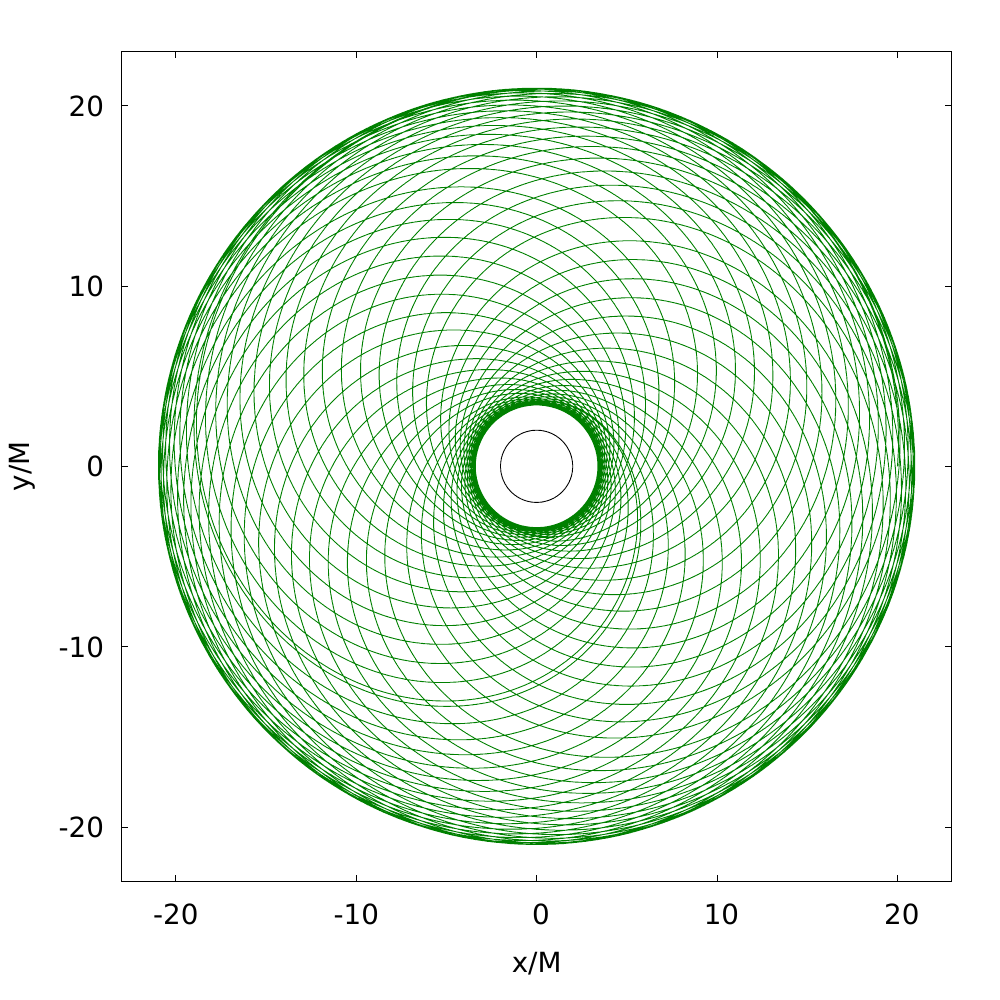}
\caption{Example of bound orbit on the equatorial plane with $\eta=5.004$ and $BM=0.1$. The orbit is bounded between $r=3.396M$ and $r=20.963M$. The central circle represents the BH {apparent} horizon.}
\label{Orbit_equatorial}
\end{figure}

Considering now the SMBH spacetime, from the LR conditions, Eqs.~\eqref{LR1} and \eqref{LR2}, we find that LRs must satisfy both $\theta=\pi/2$ and
\begin{align}
\label{LR_eq}3B^2r^3-5MB^2r^2-4r+12M=0 \ .
\end{align}
In Fig.~\ref{Veff_equatorial} (bottom panel), the effective potential {$H(r,\pi/2)$} is plotted. It may present two local extrema, corresponding to one stable and one unstable LR, which are determined by Eq.~\eqref{LR_eq}. The red dots correspond to unstable LRs, while the black {crosses} correspond to stable LRs. {In contrast to the Melvin spacetime, the SMBH does not admit a translation symmetry along the $z$ direction, so that the LRs are located on the equatorial plane only.} 
Due to the shape of the potential, bound equatorial photon orbits are allowed around the stable LR. Such orbits neither fall into the BH nor escape to infinity. In Fig.~\ref{Orbit_equatorial}, we show an example of such a bound orbit in the SMBH spacetime with $BM=0.1$.

Bound orbits are also allowed \textit{outside} the equatorial plane in the SMBH spacetime, due to the presence of \textit{closed pockets} on the effective potential {$H(r,\theta)$}, see $e.g.$~\cite{Cunha:2016bjh}. The existence of such closed pockets gives rise to interesting lensing features as we shall see below.  

As the value of $B$ is increased (in terms of the scale $M$), stable and unstable LRs converge in their radial coordinate — see the sequence of ``crosses'' in Fig.~\ref{Veff_equatorial}. For a critical value of the magnetic field strength ($B_c$) the radial coordinates of the stable and unstable LRs coincide. The value of $B_c$ is determined by Eq.~\eqref{LR_eq} together with
{
\begin{equation}
\partial^2_rH(r,\pi/2)=0,
\end{equation}
}
which has the solution
\begin{eqnarray}
B_cM = \frac{2}{5}\sqrt{\frac{169-38\sqrt{19}}{15}}\approx0.189366.
\label{Bc}
\end{eqnarray} 
For $B>B_c$, Eq.~\eqref{LR_eq} possesses no real solution with $r>2M$, and therefore there are no LRs outside the {apparent} horizon. This intriguing fact is further explored in the next section. In the following, SMBHs with $B>B_c$ shall be called \textit{overcritical}.

\subsection{{Topological charge of LRs in the SMBH and Melvin spacetimes}}
%
{It was recently shown that a four-dimensional BH spacetime that is stationary, axi-symmetric, circular and asymptotically flat must have one standard LR outside the event horizon for each rotation sense~\cite{Cunha::2020}. Subsequently, a generalization of this result for asymptotically de Sitter/anti-de Sitter spacetimes was also reported, with a similar conclusion~\cite{Wei::2020}. Since the SMBH solution is asymptotically Melvin, the existence of LRs is not guaranteed  by the above theorems; thus, the intriguing possibility of a BH without LRs emerges. As seen in the previous section, this is indeed the case for overcritical SMBHs.

In order to have a deeper understanding on the absence of LRs in a BH spacetime, it is insightful to analyze how the different (Melvin) asymptotic structure impacts on the LR topological charge, which is at the heart of the aforementioned theorems~\cite{Cunha::2017,Cunha::2020}. To do so, let us introduce the vector field $\textbf{v}=\left(\text{v}_r, \text{v}_\theta\right)$, where
\begin{align}
&\text{v}_i\equiv\frac{\partial_i H}{\sqrt{g_{ii}}}, \ \ \ \ \ \ \ \ \ i=\left(r,\theta\right) \ \ \ \ \ \text{(not summed)}.
\end{align}
Thus
\begin{align}
\label{v_r}&\text{v}_r=\frac{\sin\theta}{4r^3}\left[B^2r^2\left(3r-5M\right)-4\left(r-3M\right)\csc^2\theta\right], \\
\label{v_theta}&\text{v}_\theta=\frac{\sqrt{1-\frac{2M}{r}}\cos\theta}{16r^2\Lambda}\left[B^2r^2\left(3B^2r^2\sin^2\theta+8\right)-16\csc^2\theta\right].
\end{align}
We write the components of $\textbf{v}$ in terms of its ``norm'' $v$ and the angle $\Omega$:
\begin{align}
\label{v_r1}&\text{v}_r=\text{v}\,\cos\Omega,\\
\label{v_theta1}&\text{v}_\theta=\text{v}\,\sin\Omega,
\end{align}
where
\begin{align}
\text{v}^2\equiv \partial^\mu H\partial_\mu H=\text{v}_r^2+\text{v}_\theta^2.
\end{align}}
\begin{figure}
\centering
\subfigure{\includegraphics[scale=0.4]{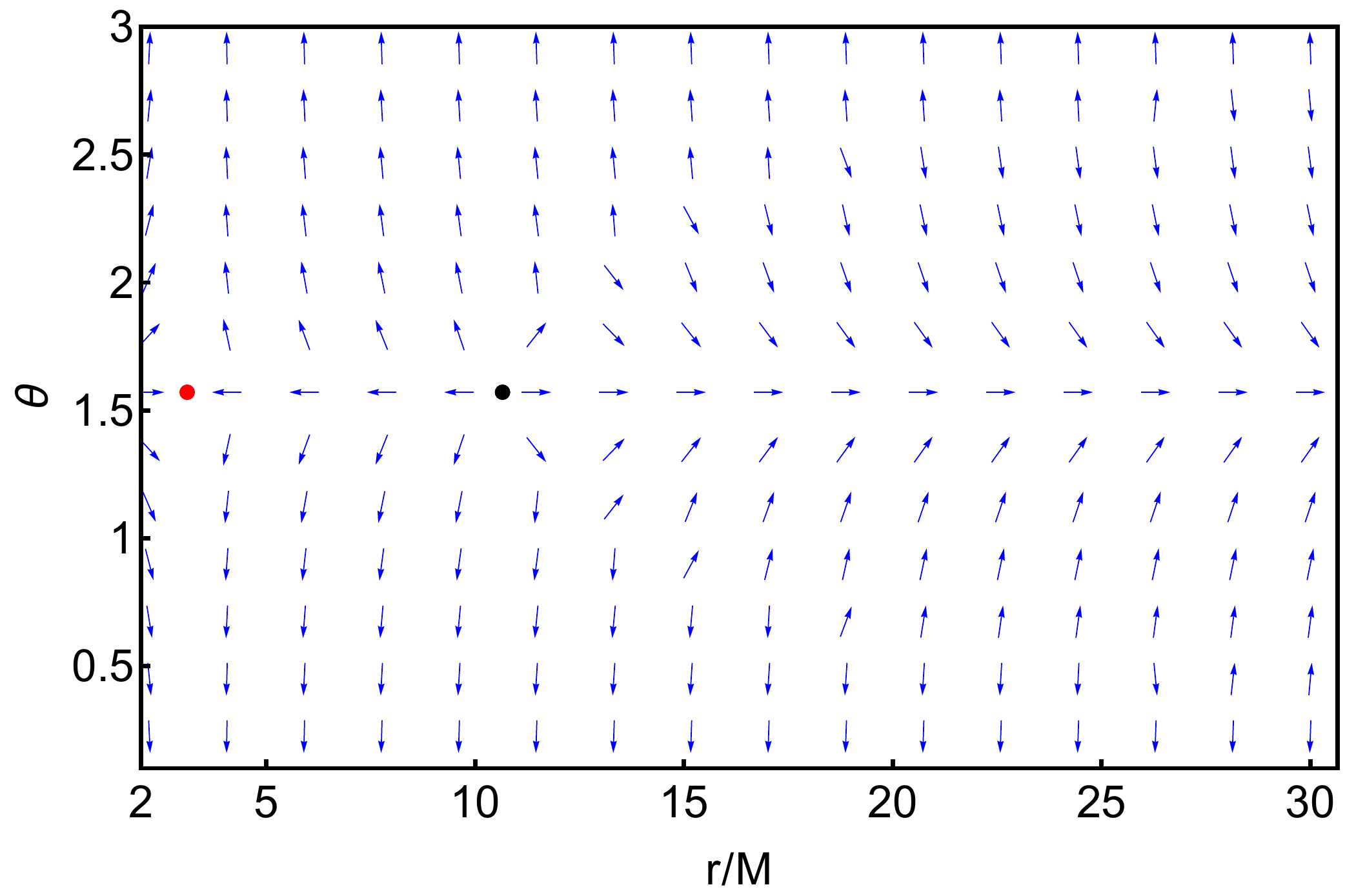}}
\subfigure{\includegraphics[scale=1.4]{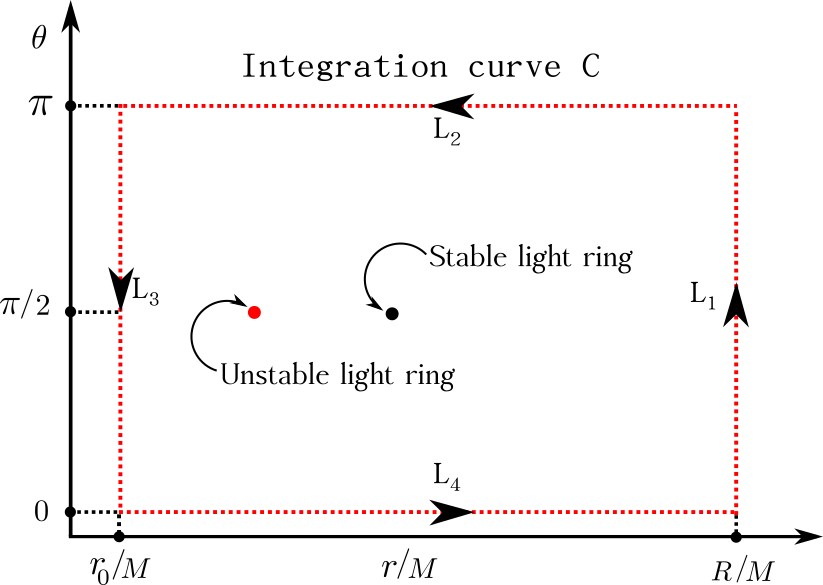}}
\caption{{Top panel: plot of the normalized vector field $\textbf{v}$ on the $(r,\theta)$ space for the SMBH spacetime with $BM=0.1$. Bottom panel: depiction of a closed (dotted) curve $C$ that encloses two LRs in the SMBH spacetime. The curve $C$ is made to approach the axis and horizon under appropriate limits (see text and~\cite{Cunha::2020} for details). }}
\label{vector_field_Ernst}
\end{figure}
{As an illustration of this vector field on the SMBH spacetime, we have represented in Fig.~\ref{vector_field_Ernst} (top panel) the field $\textbf{v}$ on the $(r,\theta)$ space. When compared to the Schwarzschild spacetime, the main difference occurs in the asymptotic region (see Fig.~1 of Ref.~\cite{Cunha::2020} for the Schwarzschild case). The Melvin asymptotics make $\text{v}_r$ asymptotically positive, whereas it is negative for asymptotically flat spacetimes. In contrast, the qualitative behavior of the vector field next to the horizon is similar for the SMBH and the vacuum Schwarzschild BH. The asymptotic difference will impact  decisively on the topological charge.
}

{To compute  the topological charge, consider a piecewise smooth and positive oriented curve $C$ in $(r,\theta)$ space; it is possible to introduce the useful integer quantity~\cite{Cunha::2020}:
\begin{equation}
\label{winding-number-int}w=\frac{1}{2\pi}\oint_C d\Omega. 
\end{equation}
If one deforms the curve $C$ without it intersecting any LR, the value of $w$ remains unaltered. Due to this, $w$ can be interpreted as a {\it topological charge} associated to $C$. Given a single nondegenerate LR that is completely inside the curve $C$, then $w$ is the topological charge of that LR~\cite{Cunha::2020}. Following~\cite{Cunha::2020}, a LR with $w=-1$ is named \textit{standard}, while a LR with $w=1$ is named \textit{exotic}~\cite{Cunha::2017,Cunha::2020}. The LR stability is actually closely connected to this value: the LR is stable (unstable) if $w=+1$ ($w=-1$), provided that the null energy condition is satisfied~\cite{Cunha::2017}, which is the case in electrovacuum.
}

{
The total topological charge of the SMBH spacetime ($M\neq 0$) is computed by choosing a curve $C$ that encloses all of the spacetime outside the BH, under some appropriate limit~\cite{Cunha::2020}. Such a possible curve $C = L_1 \cup L_2 \cup L_3 \cup L_4$ is depicted in Fig.~\ref{vector_field_Ernst} (bottom panel), where $C$ is illustrated enclosing two LRs, one stable and the other one unstable. The paths $\{L_4, L_2\}$ have constant $\theta-$values, $\theta=\{\varepsilon,\,\pi-\varepsilon\}$ respectively, whereas $\{L_1,L_3\}$ have constant radial coordinates $\{r_0, R\}$, satisfying $R> r_0 > 2M$. Thus, \eqref{winding-number-int} can be separated into the sum:
\begin{equation}
w=\frac{1}{2\pi}\left(\Delta\Omega_1+\Delta\Omega_2+\Delta\Omega_3+\Delta\Omega_4\right), \label{eqw-sum}
\end{equation}
where
\begin{align}
&\label{I1}\Delta\Omega_1=\left.\int_{L_1}\frac{d\Omega}{d\theta}d\theta\right|_{r=R},\\
&\label{I2}\Delta\Omega_2=\left.\int_{L_2}\frac{d\Omega}{dr}dr\right|_{\theta=\pi-\varepsilon},\\
&\label{I3}\Delta\Omega_3=\left.\int_{L_3}\frac{d\Omega}{d\theta}d\theta\right|_{r=r_0},\\
&\label{I4}\Delta\Omega_4=\left.\int_{L_4}\frac{d\Omega}{dr}dr\right|_{\theta=\varepsilon}.
\end{align}

The LR charge outside the BH is then obtained by taking $\lim_{R\to \infty}\,\lim_{r_0\to 2M}\,\lim_{\varepsilon\to 0} w$.
Computing analytically the integrations~\eqref{I1}-\eqref{I4} for SMBHs proved to be challenging in the most general case. However, in the large magnetic field limit, $BM\gg 1$ the calculations simplify considerably. In such limit we obtain:
\begin{align}
\label{v_r-BB}&\text{v}_r \simeq B^2\sin\theta\left(\frac{3r-5M}{4r}\right), \\
\label{v_theta-BB}&\text{v}_\theta\simeq \frac{3}{4}B^2\cos\theta\sqrt{1-\frac{2M}{r}}.
\end{align}
Taking also $\varepsilon \ll 1$, the integrations~\eqref{I1}-\eqref{I4} yield:
\begin{align}
&\label{I1-B}\Delta\Omega_1 \simeq -\pi -2\varepsilon\,\mathcal{F}(R) +\quad \mathcal{O}\left(\varepsilon^3\right),\\
&\label{I2-B}\Delta\Omega_2\simeq \varepsilon\Big[\mathcal{F}(R)-\mathcal{F}(r_0)\Big] + \quad \mathcal{O}\left(\varepsilon^3\right),\\
&\label{I3-B}\Delta\Omega_3\simeq \pi +2\varepsilon\,\mathcal{F}(r_0) +\quad \mathcal{O}\left(\varepsilon^3\right),\\
&\label{I4-B}\Delta\Omega_4\simeq \varepsilon\Big[\mathcal{F}(R)-\mathcal{F}(r_0)\Big] +\quad \mathcal{O}\left(\varepsilon^3\right),
\end{align}
where we have used the auxiliary function 
\[\mathcal{F}(r)=\frac{(5M-3r)}{3\sqrt{r(r-2M)}}.\]
The sum~\eqref{eqw-sum} then readily gives us $w \simeq \mathcal{O}\left(\varepsilon^3\right)$. Since the subleading term can be made much smaller than unity, and accounting for the fact that $w$ must be an integer, we then arrive at the neat result that $w=0$, in the large magnetic field limit $(BM\gg 1)$. 

Regarding the most general case, it is nevertheless straightforward to check numerically that the integrations~\eqref{I1}-\eqref{I4} also yield a vanishing topological LR charge for the SMBH spacetime if $B\neq 0$: 
\begin{equation}
\quad w_{\textrm{SMBH}}=0 \quad\qquad \textrm{\small $(B\neq 0)$}.
\end{equation}
In contrast, Schwarzschild has $w_{\textrm{Schw}}=-1$.  Since the topological charge $w$ is constant along the SMBH family with different $B\neq 0$, we may have one of the following two cases:
\begin{itemize}
\item (i) \textit{undercritical SMBHs} ($B<B_c$): A BH spacetime with one standard unstable LR ($w_1=-1$) and one stable exotic LR ($w_2=+1$), such that $w=w_1+w_2=0$.
\item (ii)  \textit{overcritical SMBHs} ($B>B_c$): A BH with no LRs at all outside the horizon ($w=0$).
\end{itemize}

Case (ii) is the only example (known to the authors) of a BH spacetime without any LR outside a regular horizon. This is possible due to the nontrivial asymptotic behavior of the SMBH (for $B\neq 0)$. The absence of LRs in the SMBH spacetime with $B>B_c$ was first discussed in~\cite{Dhurandhar::1983}.

\begin{figure}
\centering
\includegraphics[scale=0.35]{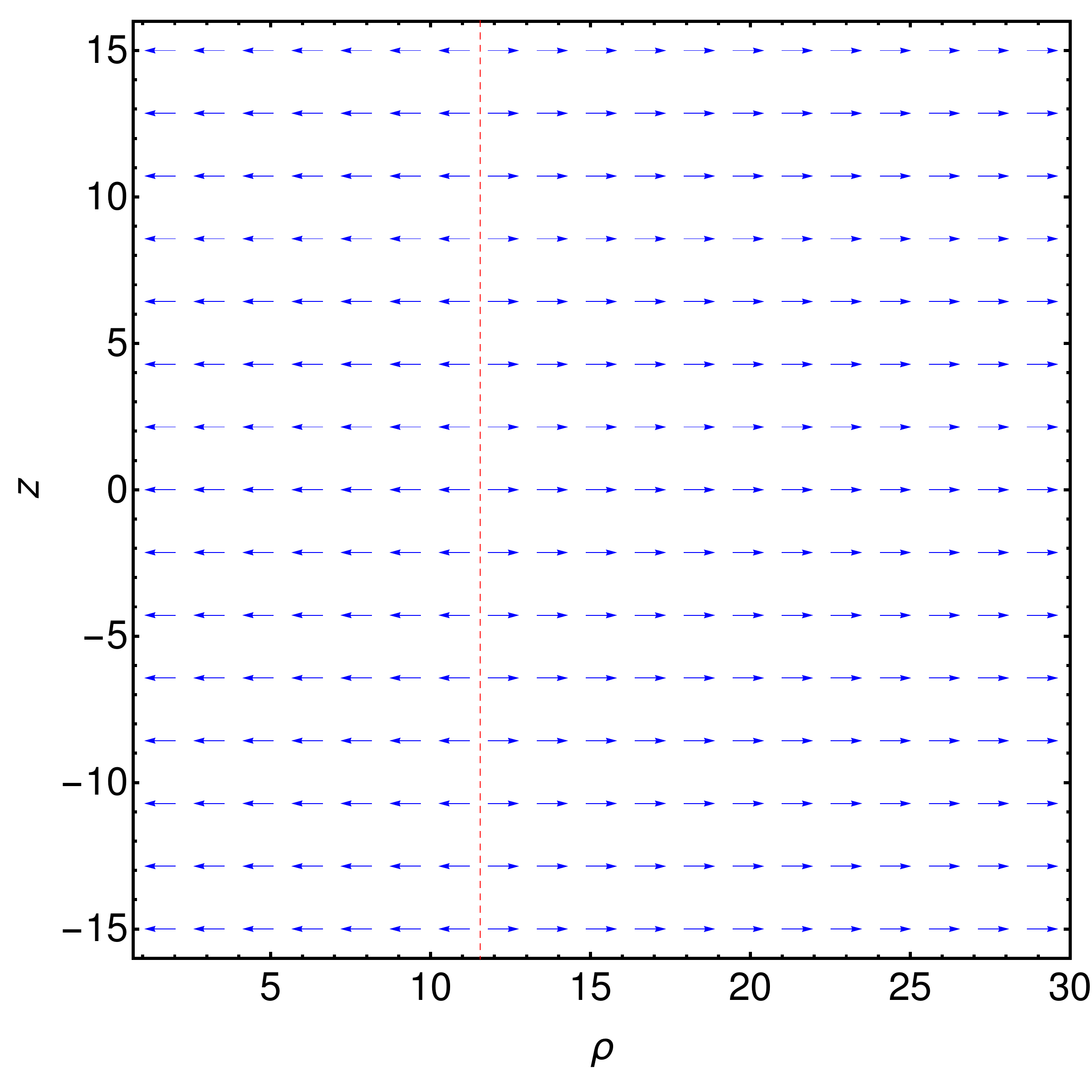}
\caption{Plot of the normalized vector field {$\textbf{v}=(\text{v}_\rho,\text{v}_z)$} in the $(\rho,z)$ plane, for the Melvin spacetime with $B=0.1$. The vertical red line represents the location of the \textit{tube of planar LRs}.}
\label{vector_field_Melvin}
\end{figure}

As a final remark concerning the analysis in this section, the contour integration approach to compute $w$ for the entire spacetime is not, strictly speaking, well defined for the Melvin $(M=0)$ case, since any closed curve $C$ setup to include virtually all of the Melvin spacetime must intersect a LR in the tube of planar LRs. To illustrate this, in Fig.~\ref{vector_field_Melvin} we represent the vector field $\textbf{v}$ in the $(\rho, z)$ plane for the Melvin spacetime, {described by
\begin{align}
&\text{v}_\rho=\frac{\partial_\rho H(\rho,z)}{\sqrt{g_{\rho\rho}}}=\frac{3B^2\rho^2-4}{4\rho^2},\\
&\text{v}_z=\frac{\partial_z H(\rho,z)}{\sqrt{g_{zz}}}=0.
\end{align}}
In that plot, the vertical red line represents the {\it tube of planar LRs}, which extends indefinitely along the $z$ direction. This tube creates an obstacle for the contour setup depicted in Fig.~\ref{vector_field_Ernst} (bottom). Although it might be possible to still define a topological charge for the Melvin universe, it will require a different approach which is beyond the scope of this paper.

\section{Shadows and gravitational lensing}
\label{Shadows}

\subsection{Setup}

Let us now turn to the gravitational lensing in the Ernst spacetime and the shadows cast by SMBHs. We use the backwards ray-tracing method, which consists in evolving the light rays from the observer's position, and backwards in time, until it is captured by the BH  or scattered to a celestial sphere of finite radius. For this purpose, we solve numerically the equations 
\begin{align}
\label{tdot}&\dot{t}=\frac{E}{\Lambda^2\,\left(1-\frac{2M}{r}\right)},\\
\label{phidot}&\dot{\phi}=\frac{L\,\Lambda^2}{r^2\sin^2\theta},
\end{align}
obtained from Eq.~\eqref{Ham1}, together with
\begin{align}
\label{rddot}\ddot{r}+\Gamma^r_{\ \mu\nu}\dot{x}^\mu\dot{x}^\nu=0,\\
\label{thetaddot}\ddot{\theta}+\Gamma^\theta_{\ \mu\nu}\dot{x}^\mu\dot{x}^\nu=0,
\end{align}
obtained from the geodesic equation. We need to supply initial conditions for Eqs.~\eqref{tdot}-\eqref{thetaddot}. This is achieved by computing the 4-momentum of the photon, as measured by a static observer. The tetrad basis of this static observer in the SMBH spacetime can be written as
\begin{align}
&\hat{e}^{\hat{0}}_{\ \mu}=\left(\sqrt{1-\frac{2M}{r}}\Lambda, 0, 0, 0\right),\\
&\hat{e}^{\hat{1}}_{\ \mu}=\left(0, \frac{\Lambda}{\sqrt{1-\frac{2M}{r}}}, 0, 0\right),\\
&\hat{e}^{\hat{2}}_{\ \mu}=\left(0, 0, r\,\Lambda,0 \right),\\
&\hat{e}^{\hat{3}}_{\ \mu}=\left(0, 0, 0, \frac{r\sin\theta}{\Lambda}\right),
\end{align}
obeying
\begin{equation}
\hat{e}^{\hat{a}}_{\ \mu}\hat{e}^{\hat{b}}_{\ \nu}n_{\hat{a}\hat{b}}=g_{\mu\nu},
\end{equation}
where $n_{\hat{a}\hat{b}}$ are the covariant components of the Minkowski metric tensor. $\hat{e}^{\hat{0}}_{\ \mu}$ is the 4-velocity of the static observer, and $\hat{e}^{\hat{i}}_{\ \mu}$ (i=1, 2, 3) are the spatial directions of the reference frame. The components of the 4-momentum of the photon measured in the static frame, namely
\begin{equation}
p^{\hat{a}}=\hat{e}^{\hat{a}}_{\ \mu}\,p^\mu,
\end{equation}
are given by
\begin{align}
\label{ptlocal}&p^{\hat{t}}=E_{obs}=\frac{E}{\sqrt{1-\frac{2M}{r}}\Lambda},\\
\label{prlocal}&p^{\hat{r}}=\frac{\Lambda}{\sqrt{1-\frac{2M}{r}}}\dot{r},\\
\label{pthetalocal}&p^{\hat{\theta}}=r\Lambda\dot{\theta},\\
\label{pphilocal}&p^{\hat{\phi}}=\frac{L\Lambda}{r\sin\theta}.
\end{align}
The photon's linear 3-momentum $\vec{p}$ in the static frame has the components
\begin{equation}
\vec{p}=\left(p^{\hat{r}}, p^{\hat{\theta}}, p^{\hat{\phi}}\right),
\end{equation}
which are parametrized as functions of the celestial coordinates $\left(\alpha, \beta\right)$ as
\begin{align}
\label{prlocal1}&p^{\hat{r}}=|\vec{p}|\cos\alpha\cos\beta,\\
\label{pthetalocal1}&p^{\hat{\theta}}=|\vec{p}|\sin\alpha,\\
\label{pphilocal1}&p^{\hat{\phi}}=|\vec{p}|\cos\alpha\sin\beta.
\end{align}
From Eqs.~\eqref{ptlocal}-\eqref{pphilocal} and Eqs.~\eqref{prlocal1}-\eqref{pphilocal1}, we obtain
\begin{align}
\label{IC0}&E=|\vec{p}|\left.\Lambda\sqrt{1-\frac{2M}{r}}\right|_{(r_{\textrm{obs}},\theta_{\textrm{obs}})},\\
\label{IC1}&\dot{r}=|\vec{p}|\Lambda^{-1}\left.\sqrt{1-\frac{2M}{r}}\cos\alpha\cos\beta\right|_{(r_{\textrm{obs}},\theta_{\textrm{obs}})},\\
\label{IC2}&\dot{\theta}=|\vec{p}|\left.\Lambda^{-1}\frac{\sin\alpha}{r}\right|_{(r_{\textrm{obs}},\theta_{\textrm{obs}})},\\
\label{IC3}&L=|\vec{p}|\left.\Lambda^{-1}r\sin\theta\cos\alpha\sin\beta\right|_{(r_{\textrm{obs}},\theta_{\textrm{obs}})},
\end{align}
where $(r_{\textrm{obs}}, \theta_{\textrm{obs}})$ are the observer's coordinates. The observer's position matches the initial photons' position, and Eqs.~\eqref{IC0}-\eqref{IC3} are the initial conditions for the equations of motion \eqref{tdot}-\eqref{thetaddot}. Each pair $(\alpha, \beta)$ represents one point in the observer's local sky. The shadow corresponds to the points in the $\alpha$~--~$\beta$ plane for which  the light rays, evolved backwards in time, using a Dormand-Prince method~\cite{C++_book}, are captured by the BH. {We compared our results, obtained with a  C\texttt{++} code with the ones obtained using the \textsc{pyhole} code~\cite{KBHSH::2016} and obtained excellent agreement.} In order to study the behavior of the scattered light rays, we place a celestial ``sphere" at $r_{cs}>r_{\textrm{obs}}$, with a  different color in each quadrant (red, green, blue, yellow). We also place a white circular spot on the celestial sphere along the axis joining the observer and the origin of the radial coordinate  — see Fig.~\ref{fig-obs} top panels. Unlike asymptotically flat localized objects, where all interesting lensing features occur near a gravitational center, in asymptotically Melvin spacetimes nontrivial features may appear \textit{panoramically}. Thus, it is of interest to consider panoramic  images, with a 360$\degree$ horizontal and 180$\degree$ vertical span.  When the BH and the magnetic field are absent ({\it i.e.} $B=0$, $M=0$),  the observer's  panoramic view is shown in Fig.~\ref{fig-obs} (bottom panel).

\begin{figure}[h!]
\centering
\includegraphics[scale=0.35]{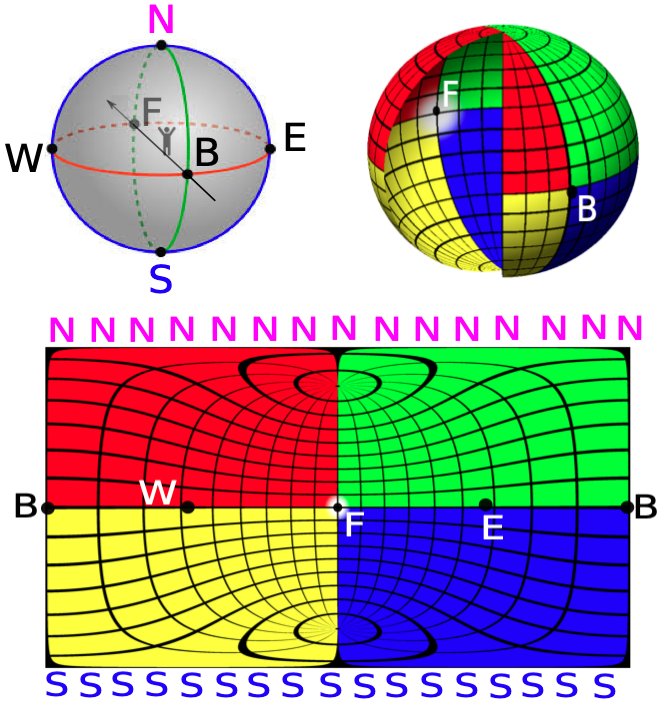}
\caption{Top left: the observer's local sky  in the equatorial position with reference directions $\{N,S,F,B,E,W\}$. Top right: the light emitting colored celestial sphere encircling both the observer and the BH. 
Bottom: the panoramic image of the observer's local sky in flat spacetime.}
\label{fig-obs}
\end{figure}

\begin{figure*}
  \centering
  \subfigure[Minkowski spacetime  ($B=0=M$), with $r_{p}=1$, $r_{cs}=2.5$ and $\theta_{\textrm{obs}}=\pi/2$.]{\includegraphics[scale=0.16]{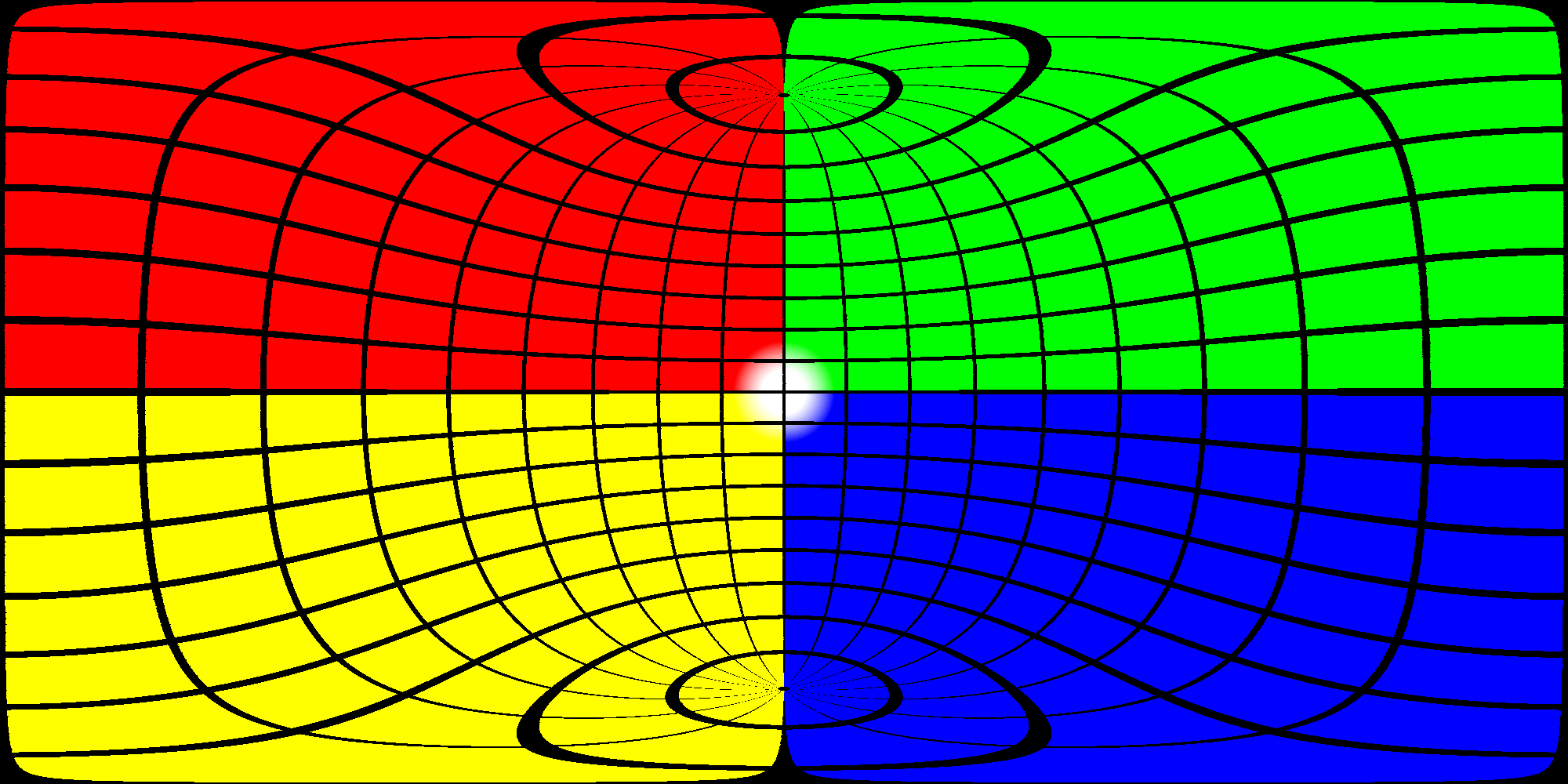}\label{a6}}\\
\subfigure[{Melvin spacetime ($B\neq 0=M$), with $B\,r_{p}=1$, $B\,r_{cs}=2.5$ and $\theta_{\textrm{obs}}=\pi/2$.}]{\includegraphics[scale=0.16]{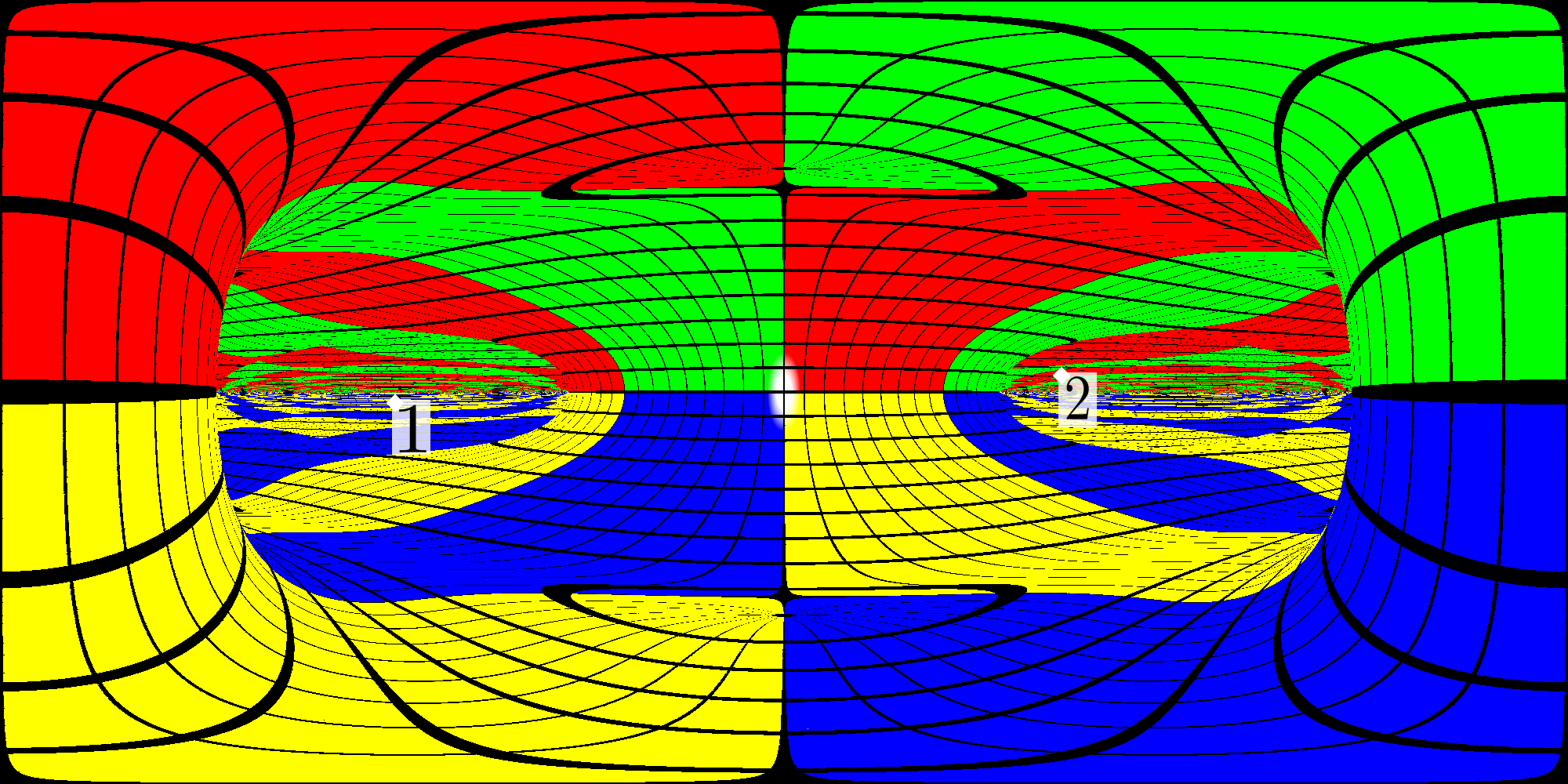}\label{Lensing_Melvin}}
\caption{Panoramic image obtained in Minkowski (upper panel) and Melvin (lower panel) spacetimes. The latter exhibits gravitational lensing. }
\label{Lensing}
\end{figure*}

In the following, SMBHs ($M,B\neq 0$) and Melvin universes ($M=0,B\neq 0$) will be said to be observed under comparable conditions if the perimetral radius $r_{p}$, given by~
\begin{equation}
r_{p}\equiv \left.\sqrt{g_{\phi\phi}}\right|_{\theta=\pi/2}=\frac{4r}{4+B^2r^2},
\end{equation}
is the same in both cases (rather than the radial coordinate $r_{\textrm{obs}}$). {In contrast to Schwarzschild~\cite{Cunha&Herdeiro::2018}, if $B\neq 0$ the perimetral radius has a maximum value, given by
\begin{equation}
\label{max-perimetral-radius}r_p^{max}=\frac{1}{B},
\end{equation}
and the corresponding value of the radial coordinate is
\begin{equation}
r=\frac{2}{B}.
\end{equation}
Since $r\in\ ]0,+\infty[$, for SMBH spacetimes with $B\neq 0$ there will be generically two radial coordinates $r$ for the same perimetral radius $r_p$. The observer location is chosen to be (of the two) the one with smallest $r$ value.
}

\subsection{{Gravitational lensing in the Melvin universe}}
\label{Subsec_Lensing_Melvin}

{The gravitational lensing in the Melvin universe ($B\neq0=M$) is exhibited in Fig.~\ref{Lensing_Melvin}, where it is compared with the Minkowski case, Fig.~\ref{a6}. For the Melvin universe, $B$ is the only scale and so all dimensionful quantities are normalized  by $B$. The panoramic image is for an observer at $r_p=1/B$, $\theta_{\textrm{obs}}=\pi/2$ and a celestial sphere at $r_{cs}=2.5/B$.

The Melvin spacetime (unlike Minkowski), exhibits gravitational lensing leading to several notable features:
\begin{description}
\item[1)] Looking forward (toward F), the observer sees the celestial sphere flipped horizontally, i.e., left-right inverted.  This left-right flipping is a consequence of the axial symmetry together with staticity. By contrast, in spherical symmetry, a gravitational centre tends to flip the image with respect to the symmetry center, rather than an axis — see, $e.g.,$ Fig. 3 (top right) in~\cite{KerrwSH_Shadow}.
\item[2)] Looking east/west (toward E/W), the observer sees some chaotic regions. An understanding of these regions can be obtained by following illustrative photons. For this purpose we have highlighted two points ({\bf 1} and {\bf 2}) of the image within the turbulent regions.
Introducing a radial coordinate that compactifies the radial direction
\begin{equation}
\mathcal{R}_M\equiv\frac{r}{1+r}.
\end{equation}
\begin{figure*}
\centering
\subfigure{\includegraphics[scale=0.46]{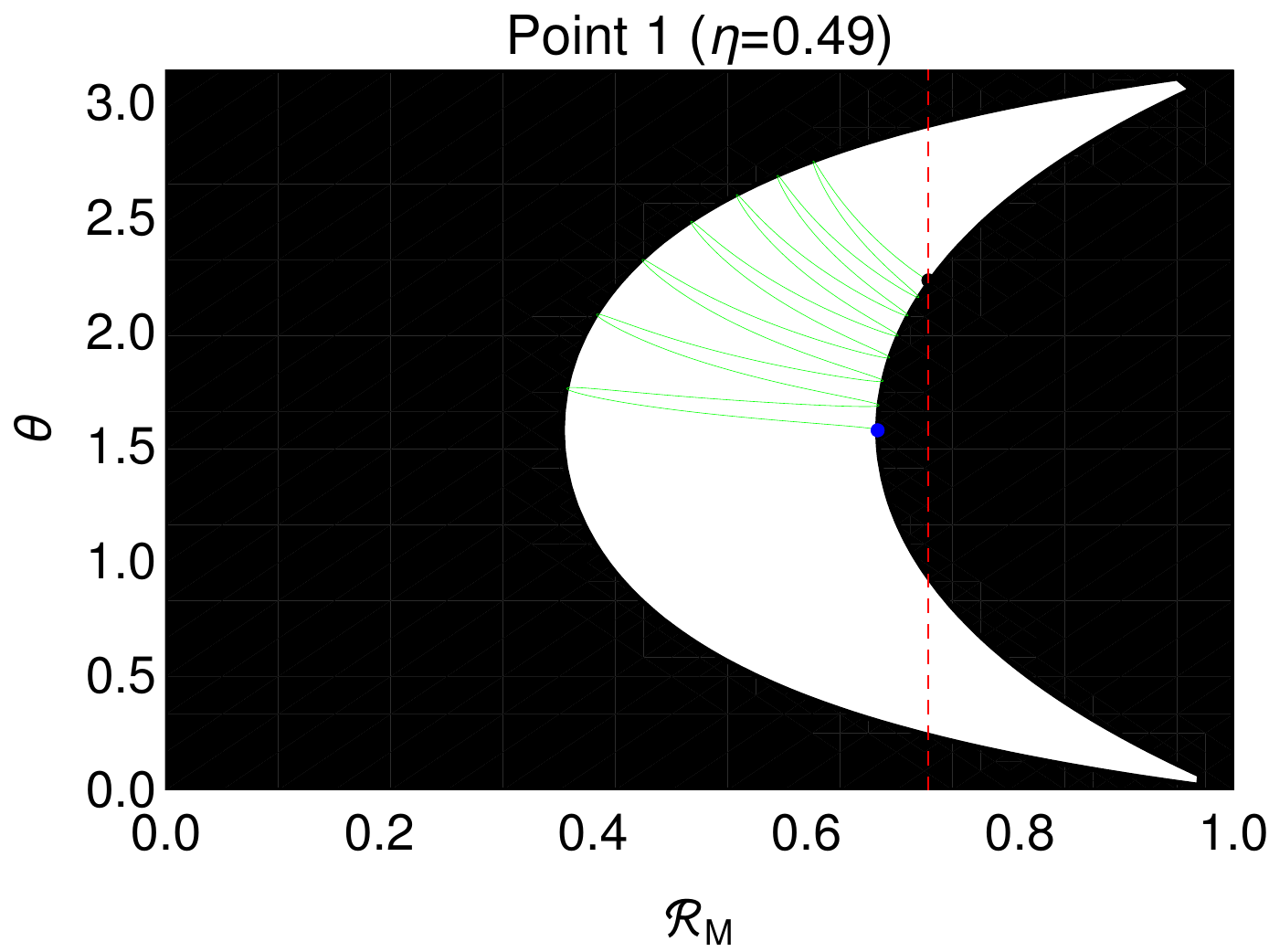}}
\subfigure{\includegraphics[scale=0.4]{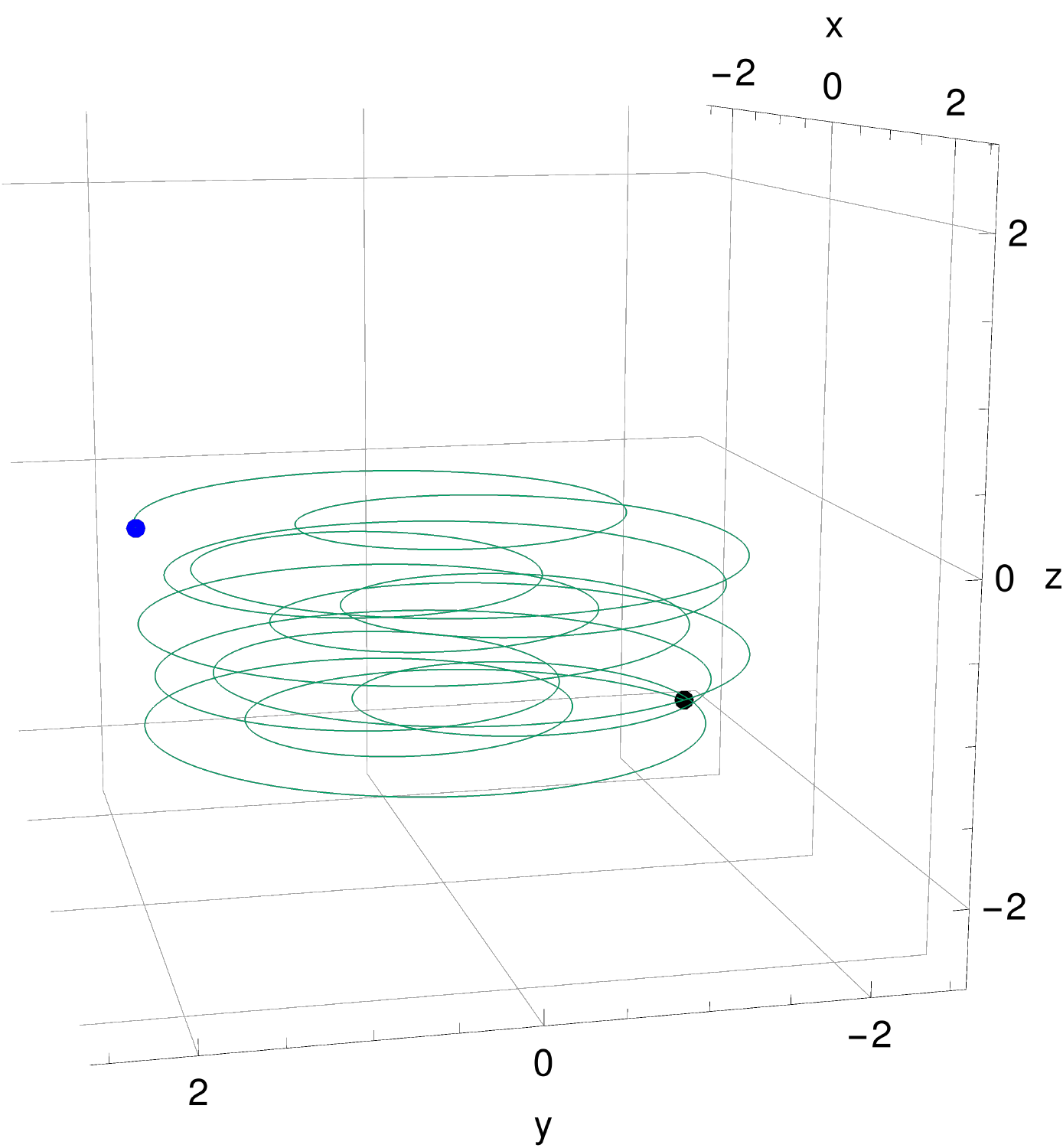}}\\
\subfigure{\includegraphics[scale=0.46]{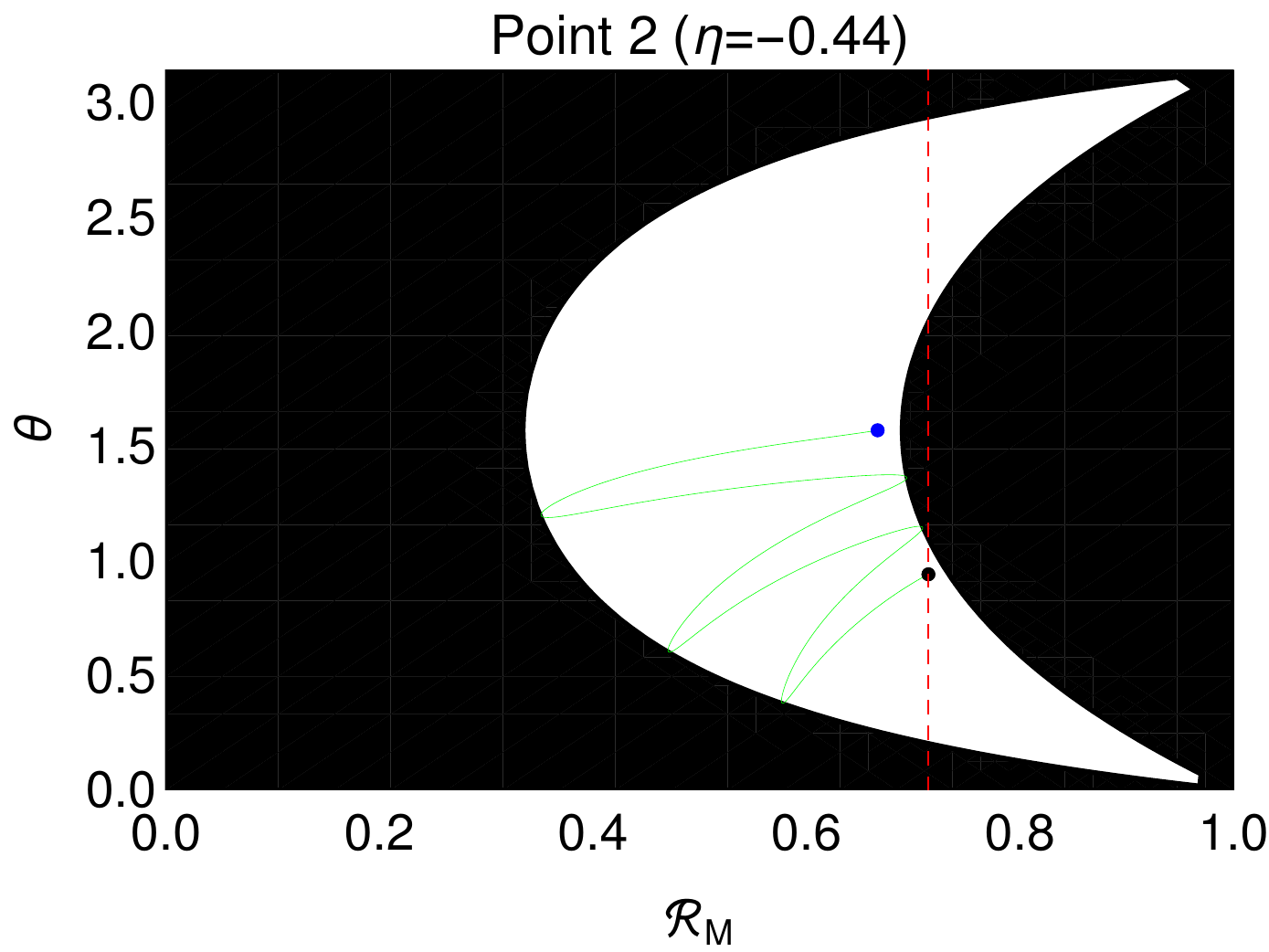}}
\subfigure{\includegraphics[scale=0.4]{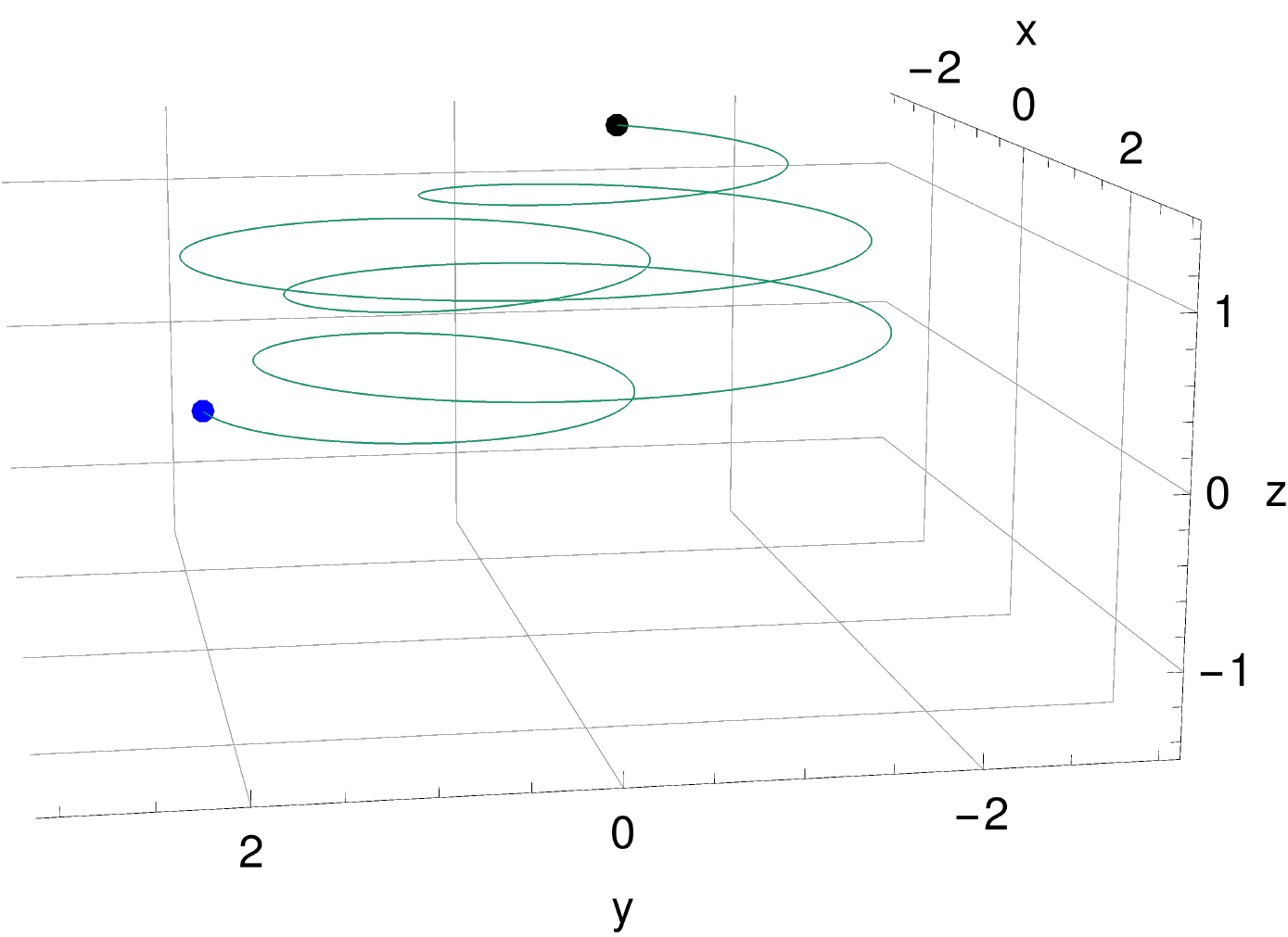}}\\
\caption{{Left: contour plots of the effective potential for the Melvin spacetime in the $\mathcal{R}_M-\theta$ plane, for the points marked in Fig.~\ref{Lensing_Melvin}. The trajectories start off at the blue dot (observer) and terminate at the black dot, where they meet the celestial sphere (red dotted vertical line). Right: trajectories described by the photon for the corresponding values of $\eta$ (displayed in the left column).}}
\label{Veff_off_equatorial_melvin}
\end{figure*}
Figure~\ref{Veff_off_equatorial_melvin}  exhibits the effective potential $H(r,\theta)$ contour plot, in terms of  $\mathcal{R}_M$, for the Melvin universe and the photons corresponding to the points highlighted in Fig.~\ref{Lensing_Melvin}. The black regions in the left panels of Fig.~\ref{Veff_off_equatorial_melvin}  are the inaccessible regions for the photon due to the restriction~\eqref{geo_cond} on the effective potential. Recall that spatial infinity is inaccessible to these photons (or any photons that have $\eta\neq 0$). The photons start off at the blue dot (observer) and terminate their trajectories at the black dot, where they meet the celestial sphere (red dotted line). The left panels of Fig.~\ref{Veff_off_equatorial_melvin}, clarify that the turbulent lensing region arises due to the existence of multiple turning points of the orbit before encountering the celestial sphere, which occurs at larger latitudes than that of the initial observation. 
\item[3)]  Looking north/south or back (toward N/S or B), the observer sees a qualitatively similar image as to that in Minkowski spacetime in terms of the color patterns, albeit a different structure for the grid lines, likely a consequence of $r$=constant surfaces in~\eqref{line_el} not being round spheres.
\end{description}

\subsection{Undercritical SMBHs}
\label{Subsec_shadow_Ersnt}
We now turn our attention to the general SMBH spacetime ($M\neq0$) considering not only the lensing but also the shadows produced by the existence of a BH region. Let us start by considering how the Schwarzschild shadow is deformed due to \textit{small} magnetic fields and undercritical SMBHs. For this purpose, we shall consider \textit{nonpanoramic, narrrow angle} images, centered around the F direction.

Figure~\ref{Shadow1} exhibits the shadows and lensing of SMBHs for $BM \ll 1$, which is (likely) the parameter range most suitable to model real astrophysical environments (near the BH region). In this figure the observer is located on the equatorial plane ($\theta_{\textrm{obs}}=\pi/2$), at $r_p=10M$ and the viewing angle is $75\degree$ (as opposed to $360\degree$ in the panoramic images).
The colored celestial sphere in Fig.~\ref{Shadow1}
has radius $r_{cs}=25M$. For $B=0$, we recover the Schwarzschild BH shadow, which is perfectly circular in the observer's local sky. For $B> 0$, we note that the shadow is no longer circular, becoming \textit{oblate} (elongated along the horizontal direction) as the value of $B$ is  increased. It is a rather curious feature in the SMBH spacetime that the intrinsic horizon geometry becomes prolate (cf. Fig.\ref{Horizon_embedding}) but the BH shadow becomes oblate (cf. Fig.~\ref{Shadow1}). This is yet another manifestation of how the shadow is \textit{not} a faithful probe of the event horizon  geometry~\cite{Cunha:2018gql}.

\begin{figure*}
  \centering
  \subfigure[BM=0]{\includegraphics[scale=0.1]{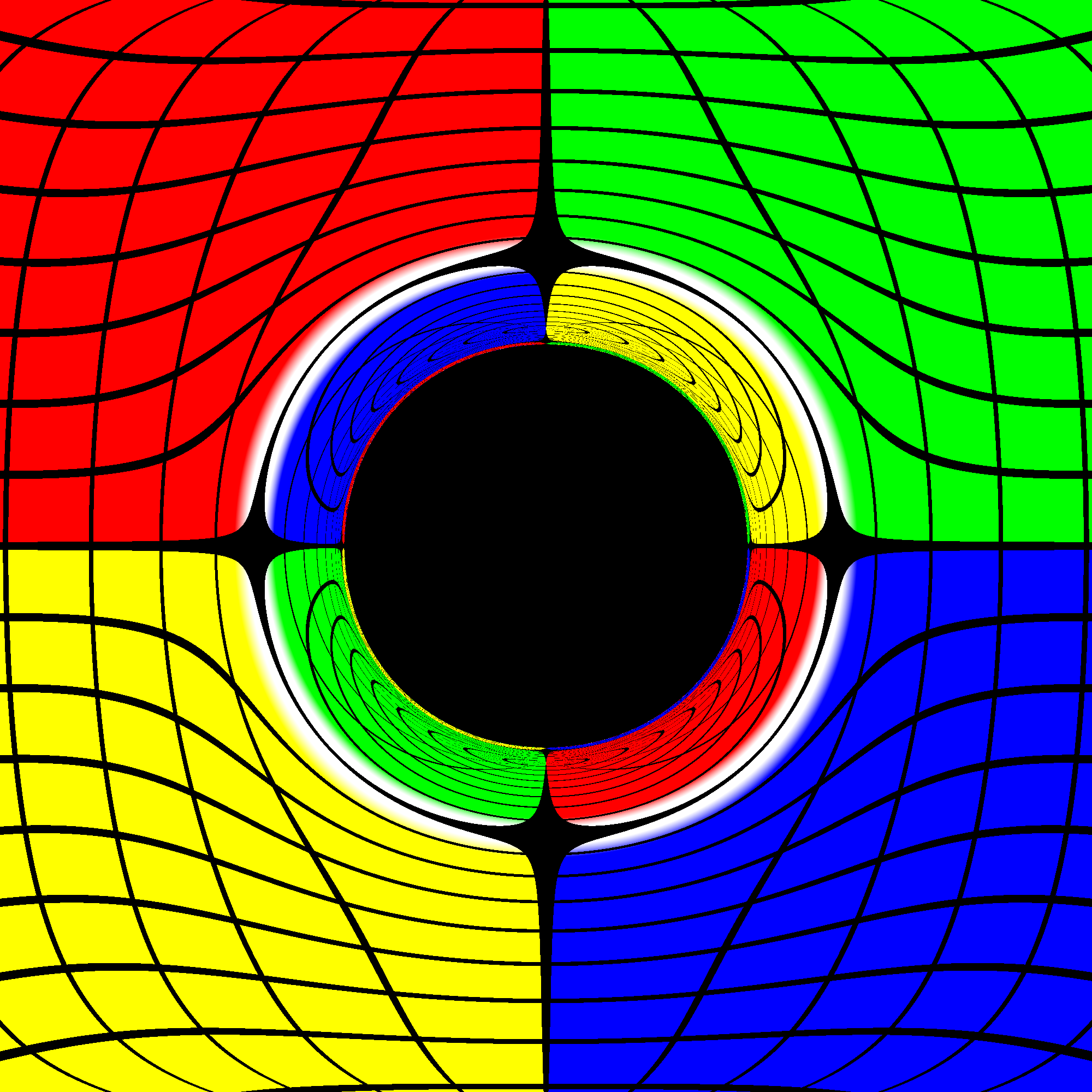}\label{a}}
  \subfigure[BM=0.03]{\includegraphics[scale=0.1]{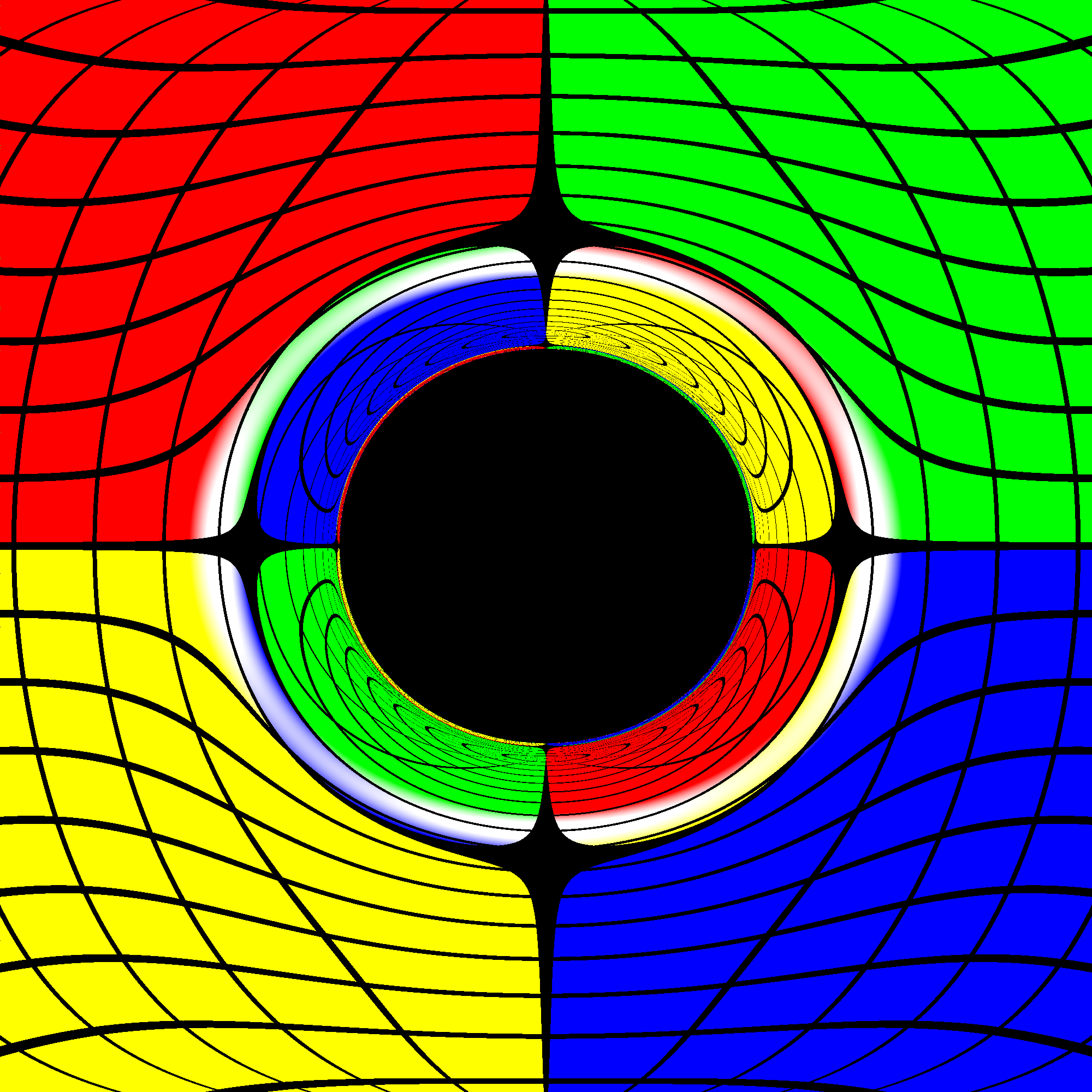}\label{b}}
  \\
  \subfigure[BM=0.04]{\includegraphics[scale=0.1]{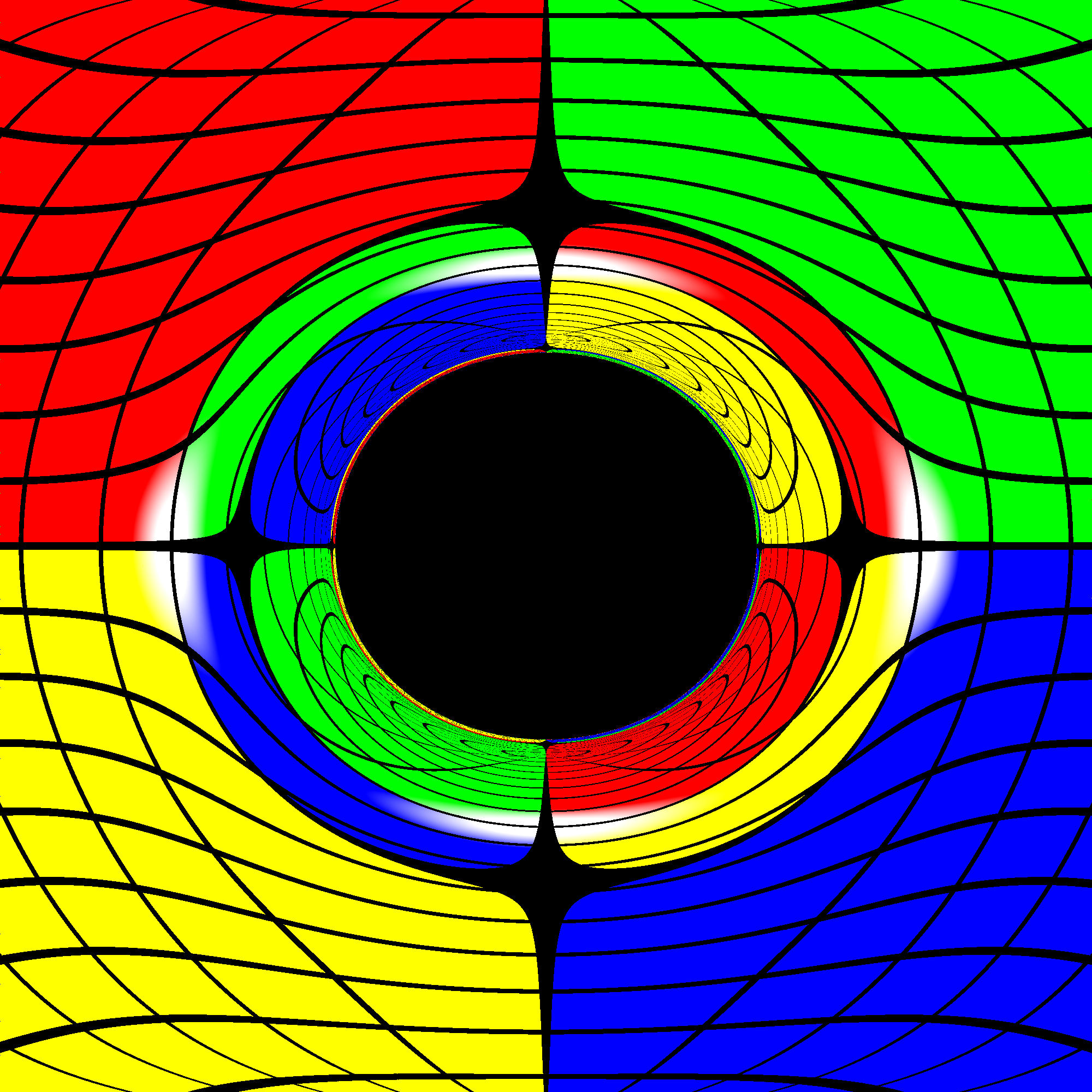}\label{c}}
    \subfigure[BM=0.05]{\includegraphics[scale=0.1]{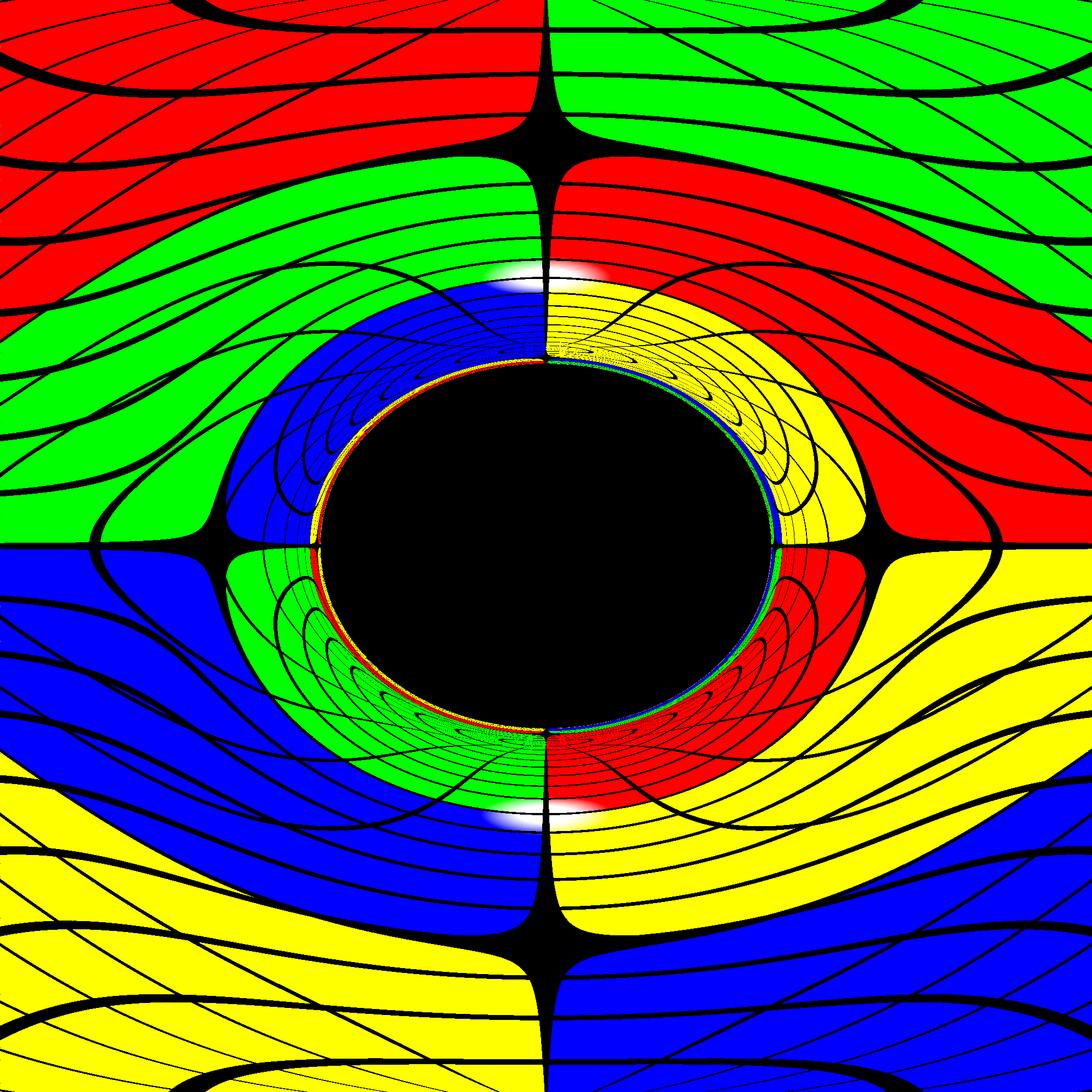}\label{d}}
\caption{Nonpanoramic images of the shadows and gravitational lensing of (undercritical) SMBHs for different values of the magnetic field $B$. We have chosen the observer perimetral radius $r_p=10M$, $\theta_{\textrm{obs}}=\pi/2$ and $r_{cs}=25M$.}
\label{Shadow1}
\end{figure*}

\begin{figure*}
\centering
\subfigure[]{\includegraphics[scale=0.2]{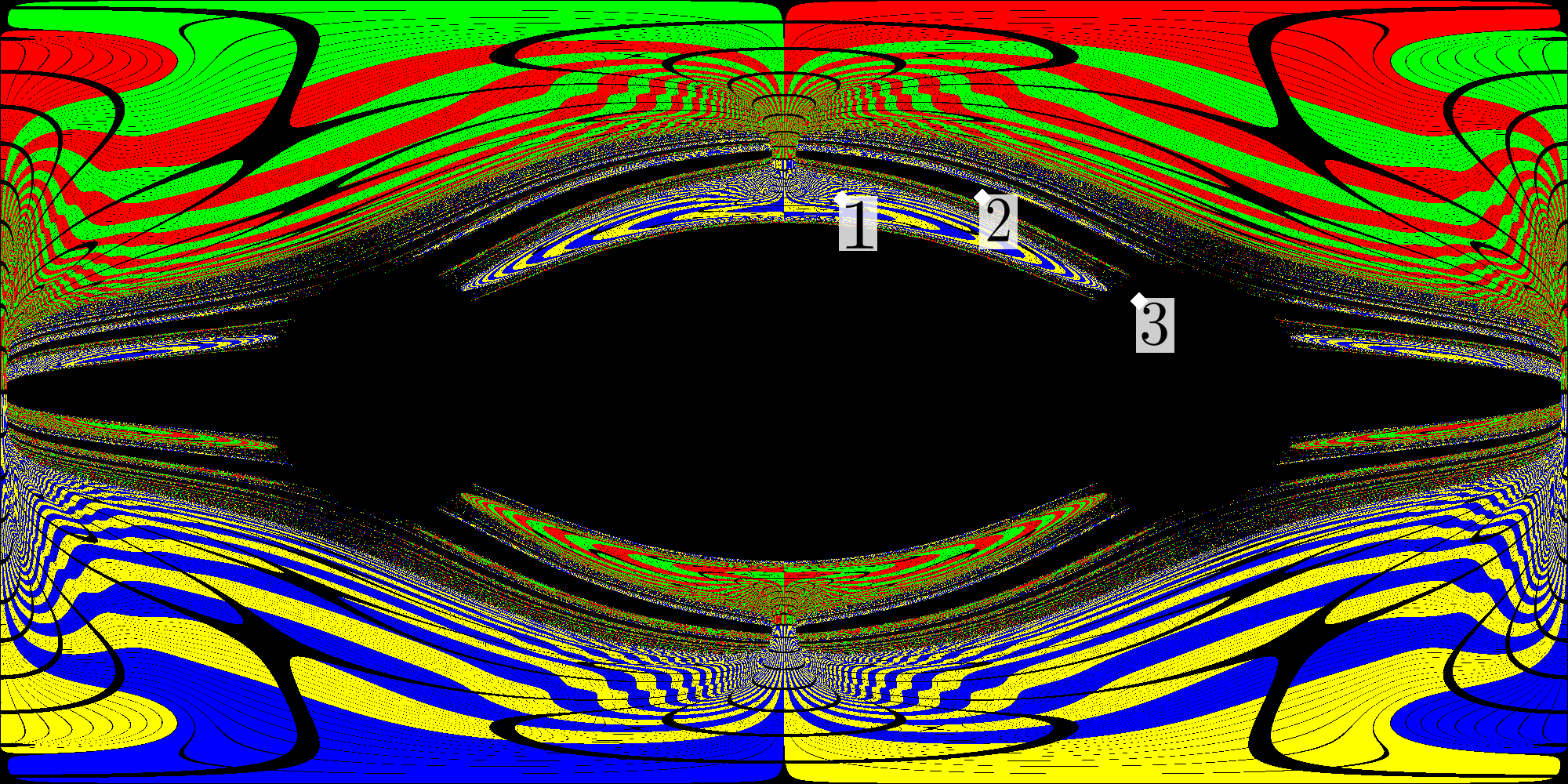}\label{ShadowB01_WP}}
\subfigure[]{\includegraphics[scale=0.2]{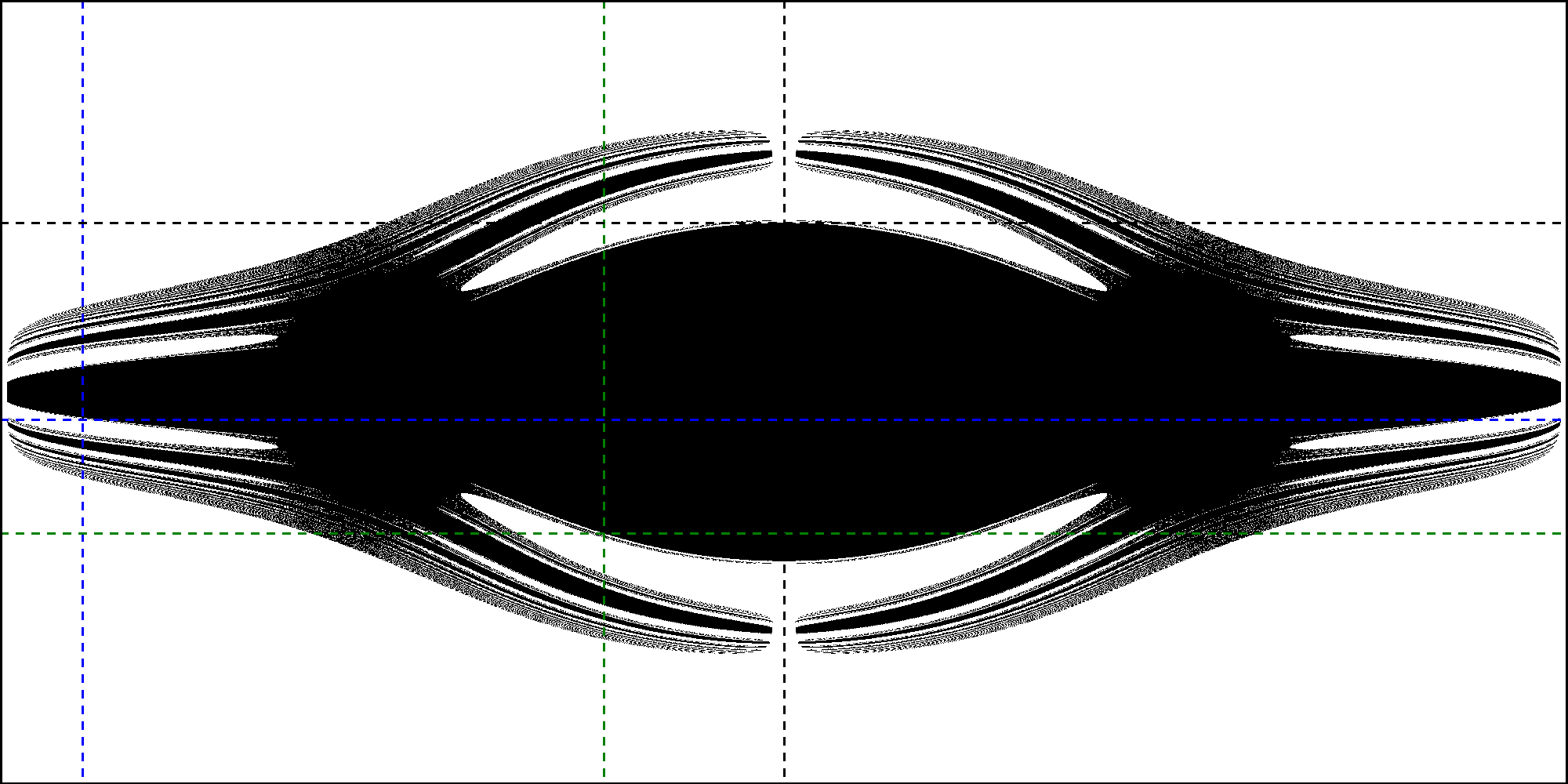}\label{ShadowB01-middle}}\\
\subfigure[]{\includegraphics[scale=0.55]{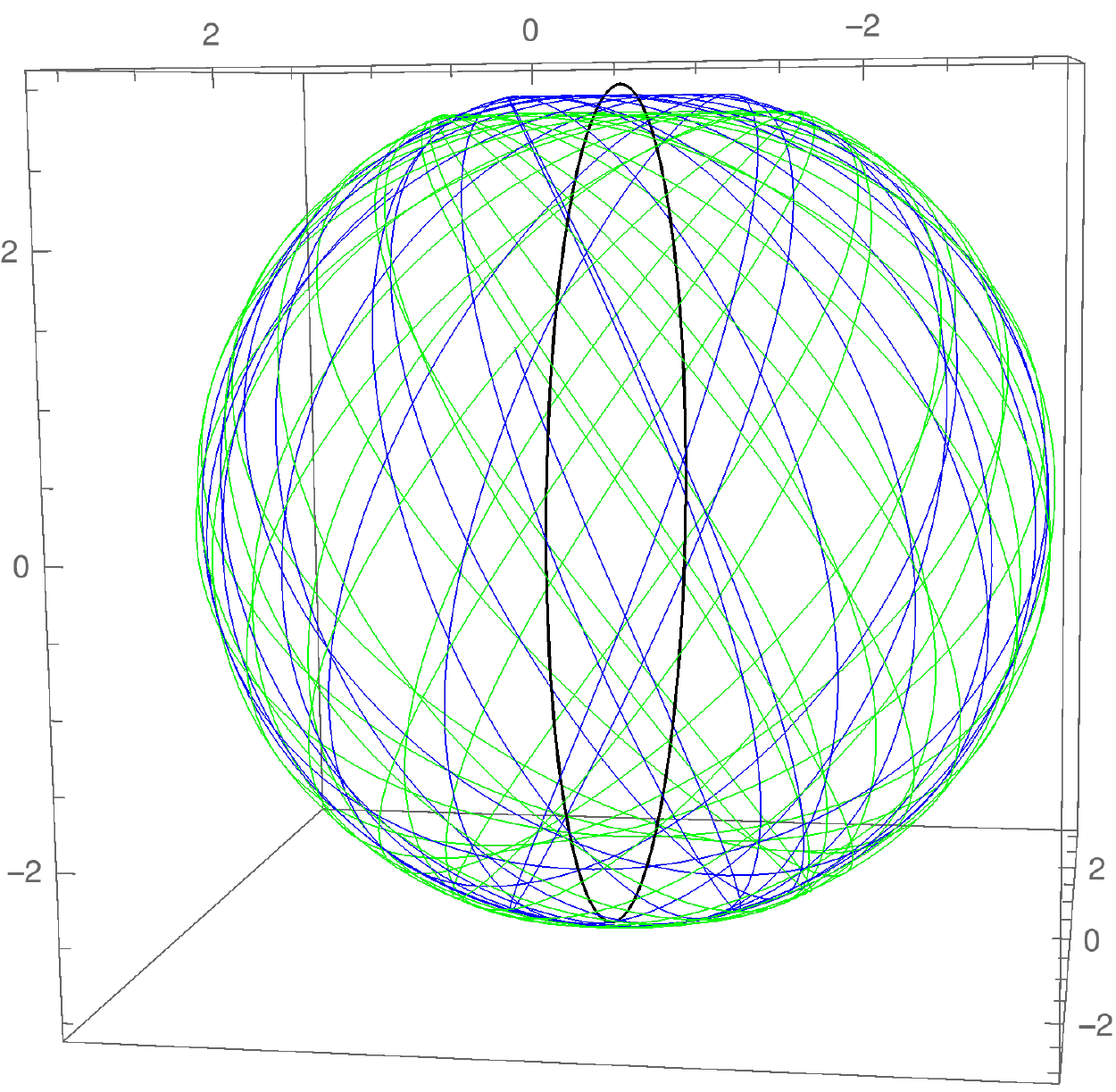}\label{FPO}}
\caption{Top: panoramic image of the shadow and gravitational lensing of an overcritical SMBH ($BM=0.2$).  The observer is located at $r_p=4.698M$, $\theta_{\textrm{obs}}=\pi/2$ and $r_{cs}=45M$. Middle: the celestial sphere was colored white to emphasize the multiple disconnected shadows (black regions) in the top image. Bottom: three FPOs (blue, green and black curves), corresponding to three points at the shadow edge in the middle panel, at the intersection  of vertical and horizontal lines with the corresponding color.
}
\label{ShadowB01}
\end{figure*}

Further analyzing the impact of B on the shadow in Fig.~\ref{Shadow1}, we remark that in the Schwarzschild case, the white spot on the celestial sphere appears as a ring in the observer's local sky: an \textit{Einstein ring}~\cite{Einstein_ring}. For the SMBH, on the other hand, the spherically symmetric Einstein ring is broken up; as the value of $B$ is increased one observes different patterns.  For instance, for $BM=0.04$ [Fig.~\ref{c}] four (distorted) copies of the white spot are seen, forming the ends of a cross centered in the BH shadow,  somewhat reminiscent of the famous Einstein cross formed by the lensed image of the quasar QSO 2237+0305.

\subsection{Overcritical SMBHs}

Let us now consider \textit{large} magnetic fields and overcritical SMBHs. As we have seen in Sec.~\ref{sec3a}, for $B>B_c$ one of the most remarkable features concerning null geodesics in the SMBH spacetime arises: there are no LRs. This leads to the following intriguing question: since LRs determine the edge of the BH shadow (on the equatorial plane), what happens to the shadow of SMBHs with $B>B_c$?

To investigate this question,  Fig.~\ref{ShadowB01} exhibits the shadow and gravitational lensing for a SMBH with $BM=0.2$ [thus $B>B_c$, cf. Eq.~\eqref{Bc}]. This is a panoramic image since the impact  of the absence  of LRs is to permit the possibility of a \textit{panoramic, (almost) $360\degree$ wide, BH shadow}.

Let us  dissect the previous statement. In Fig.~\ref{ShadowB01}, the observer's perimetral radius and the radius of the celestial sphere are $r_p=4.698M$ and $r_{cs}=45M$, respectively.  The location of the celestial sphere impacts on how panoramic the  shadow is. Figure~\ref{Veff_equatorial} makes clear that outwards directed equatorial light rays with $\eta\neq  0$ always reach a radial turning point. This  turning point is at an increasingly large $r$ when approaching the backwards (B) direction, and $r\rightarrow \infty$ as $\eta=0$, i.e.  as the observer looks toward B.  Hitting the celestial sphere is a stoppage criterion, yielding a colorful pixel. But all lights rays except the one with $\eta=0$, would reach a radial turning point whence they necessarily fall back into the BH region (green curve in  Fig.~\ref{Veff_equatorial}) if the celestial sphere is pushed to a sufficiently large $r_{cs}$, yielding a panoramic (almost)  $360\degree$ wide BH shadow.\footnote{Strictly speaking, there is always a vicinity of the B direction  (backwards), wherein no shadow is seen, since such photons hit the celestial sphere before reaching their radial turning point no matter how large (but finite) $r_{cs}$ is. But for a sufficiently far away celestial sphere this vicinity is reduced, in the discretized image, to a single point: the B direction.}

Besides the panoramic (equatorial) shadow, the image of overcritical SMBHs in Fig.~\ref{ShadowB01} has other outstanding features.
In Fig.~\ref{ShadowB01-middle} the shadow image is exhibited with a white colored celestial sphere, to appreciate the appearance of multiple disconnected shadows. Thus both the shadow and the lensing [as seen in Fig.~\ref{ShadowB01_WP}] contain imprints of a chaotic pattern.

In order to better understand these chaotic patterns in the overcritical SMBH we proceed as in the Melvin universe case, analysing individual orbits. In Fig.~\ref{ShadowB01_WP} we have highlighted three distinct points on the celestial sphere: one scattered orbit in the chaotic region (point {\bf 1}) and two absorbed orbits in two distinct shadow regions (points {\bf 2} and {\bf  3}). The analysis of the corresponding  orbits is done in Fig.~\ref{Veff_off_equatorial}, specifying the value of $\eta$ and the contour plots of the effective potential $H(r,\theta)$. As in Fig.~\ref{Veff_off_equatorial_melvin}, the blue (black) dots mark the beginning (end) of the orbit, corresponding to the position of the observer  (celestial sphere or apparent horizon). In the right panels of Fig.~\ref{Veff_off_equatorial}, the black sphere has radius equal to the one of the BH {apparent} horizon. In the left panels of Fig.~\ref{Veff_off_equatorial},  the radial coordinate along the horizontal axis is the compactified coordinate~\cite{KBHSH::2016}
\begin{eqnarray}
\mathcal{R}\equiv\frac{\sqrt{r^2-(2M)^2}}{1+\sqrt{r^2-(2M)^2}},
\end{eqnarray}
such that spatial infinity corresponds to $\mathcal{R}=1$, and the {apparent} horizon to $\mathcal{R}=0$.

 \begin{figure*}
  \subfigure{\includegraphics[scale=0.5]{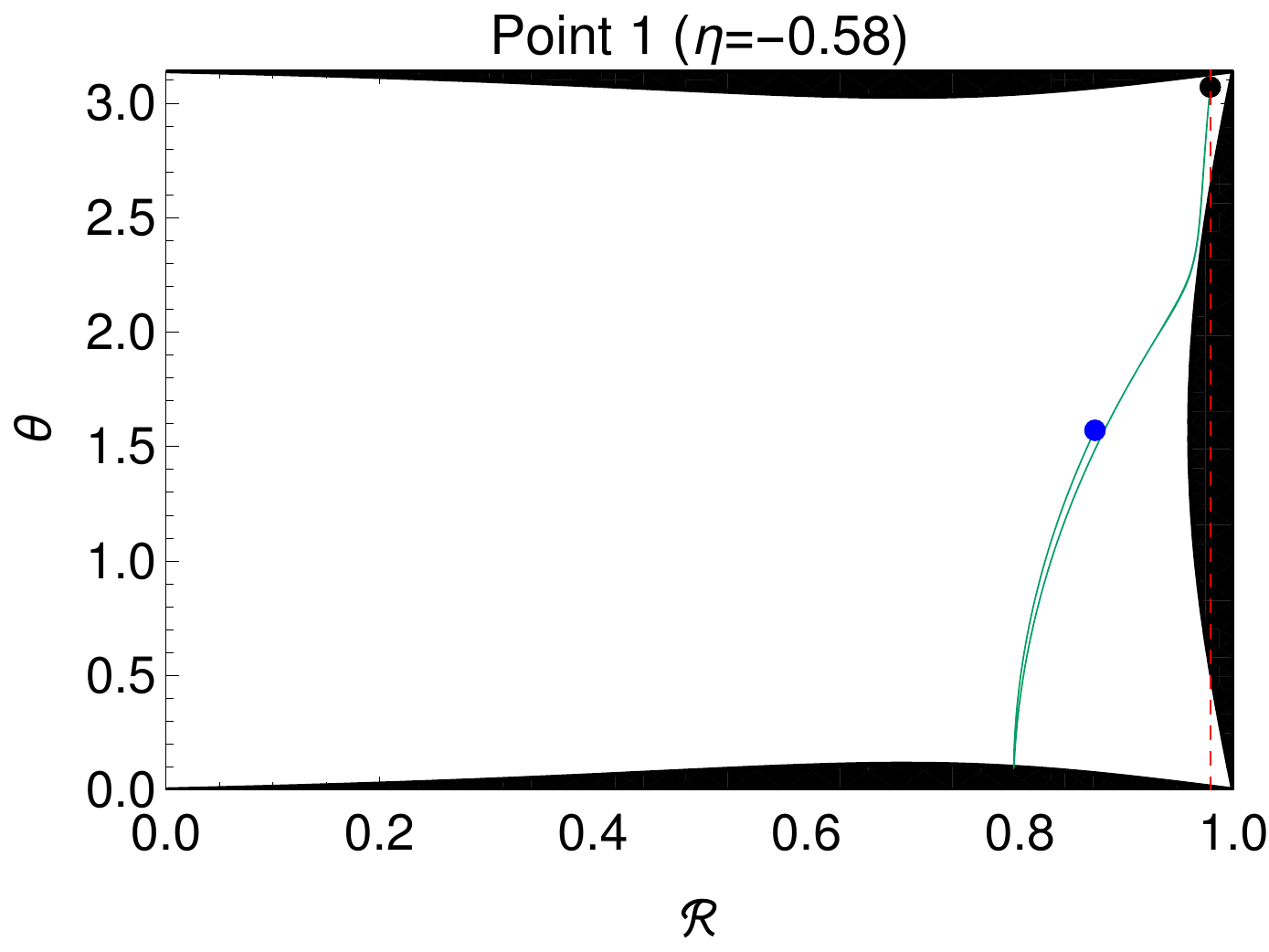}}
  \subfigure{\includegraphics[scale=0.4]{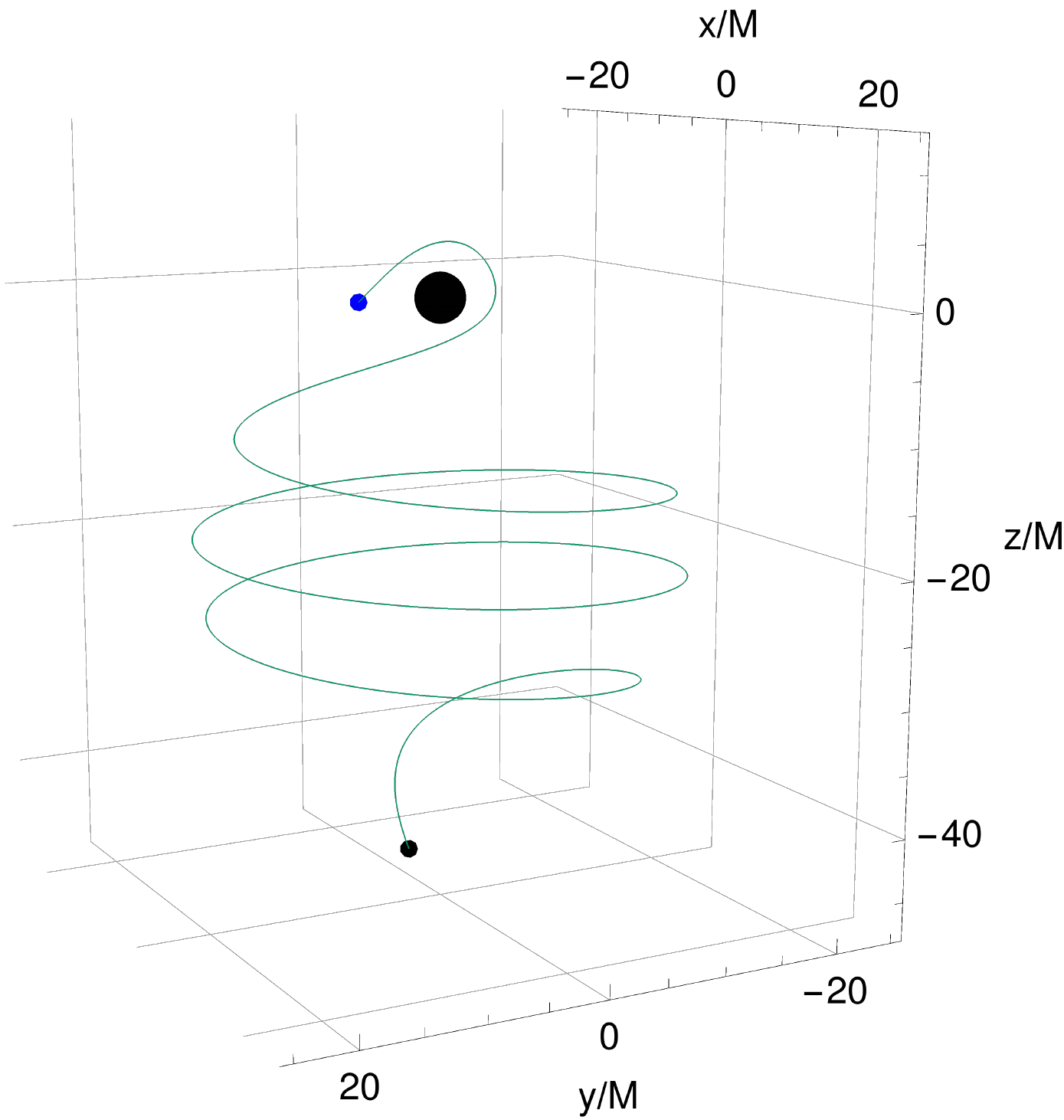}}\\
    \subfigure{\includegraphics[scale=0.5]{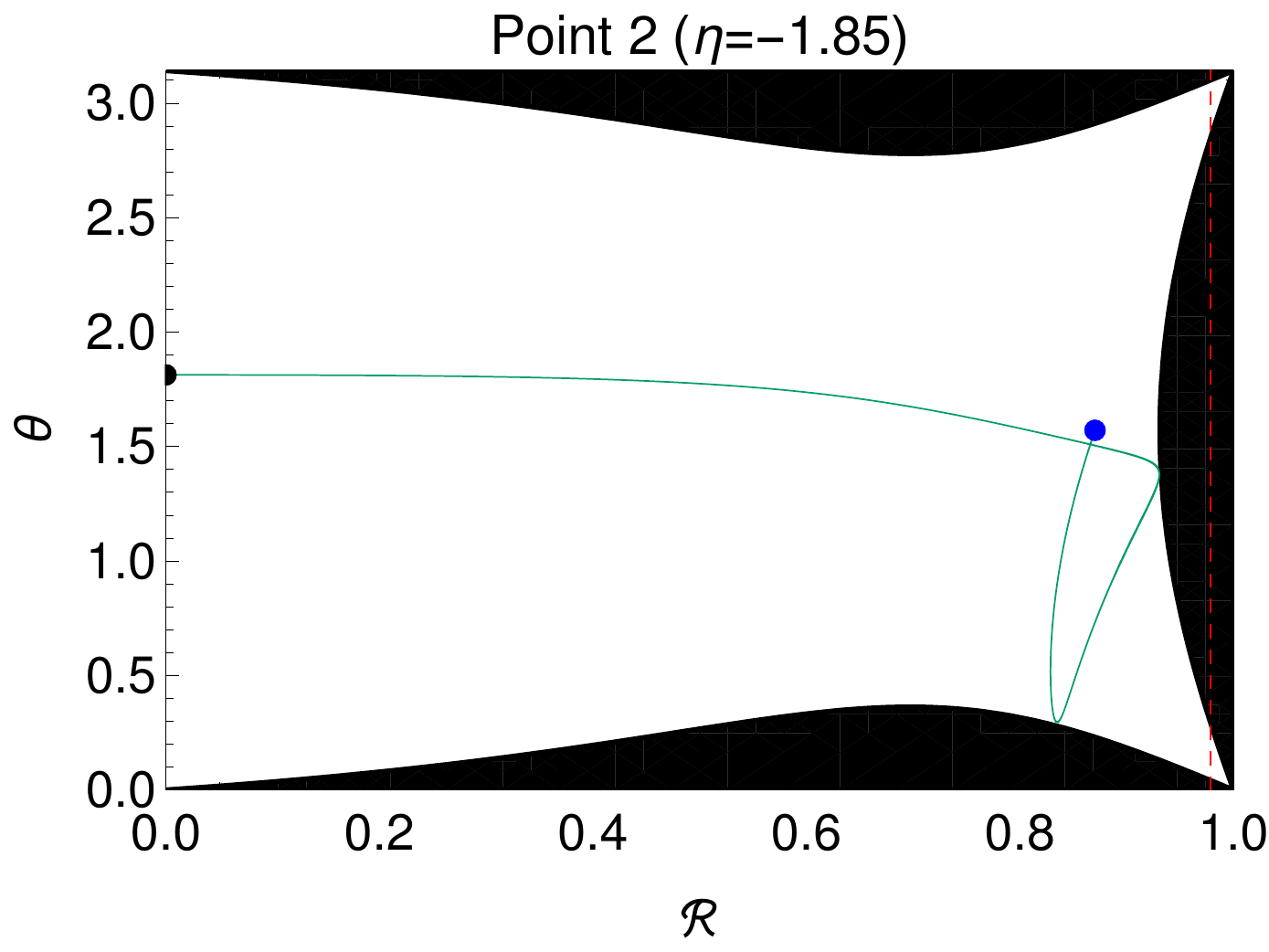}}
  \subfigure{\includegraphics[scale=0.41]{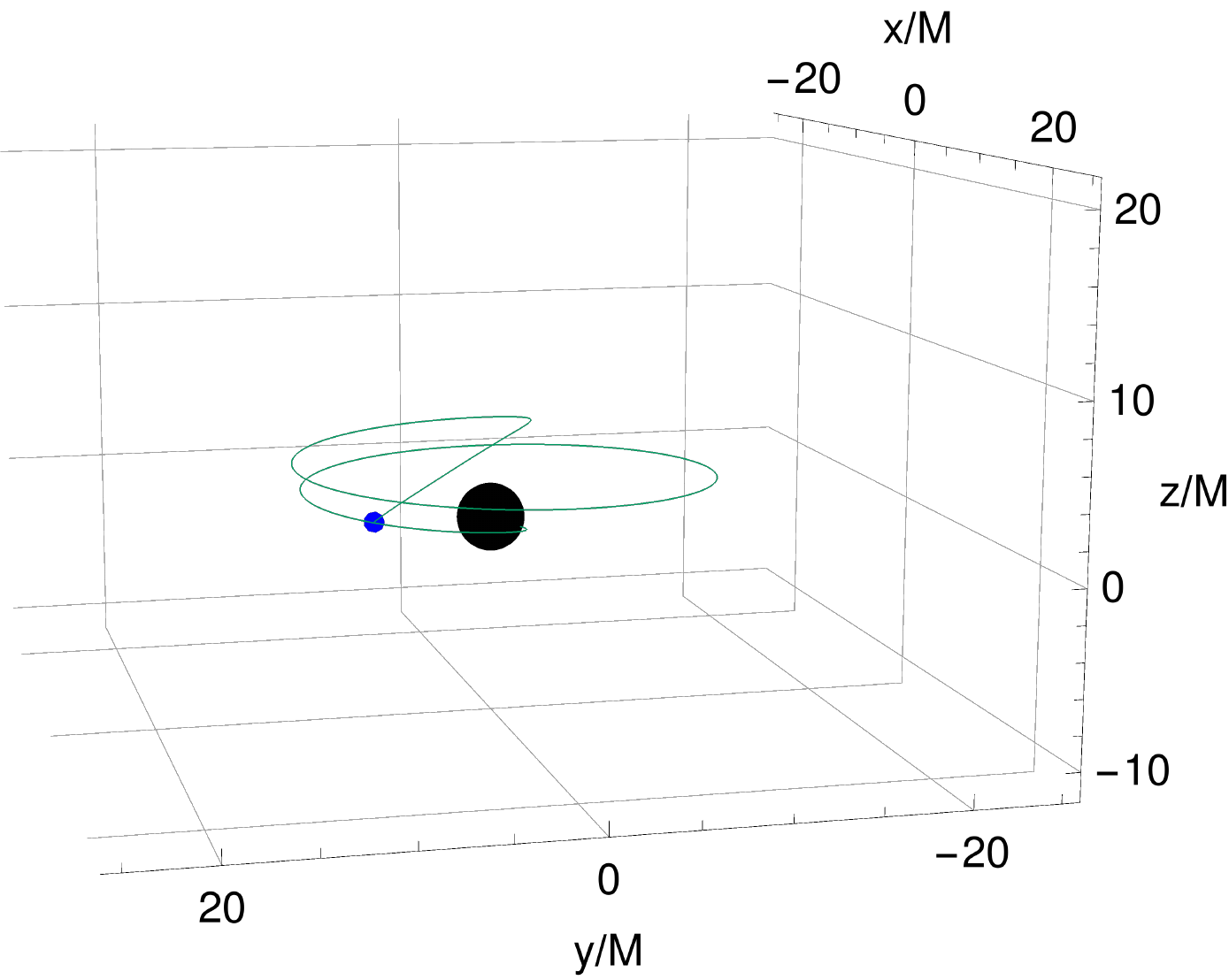}}\\
      \subfigure{\includegraphics[scale=0.5]{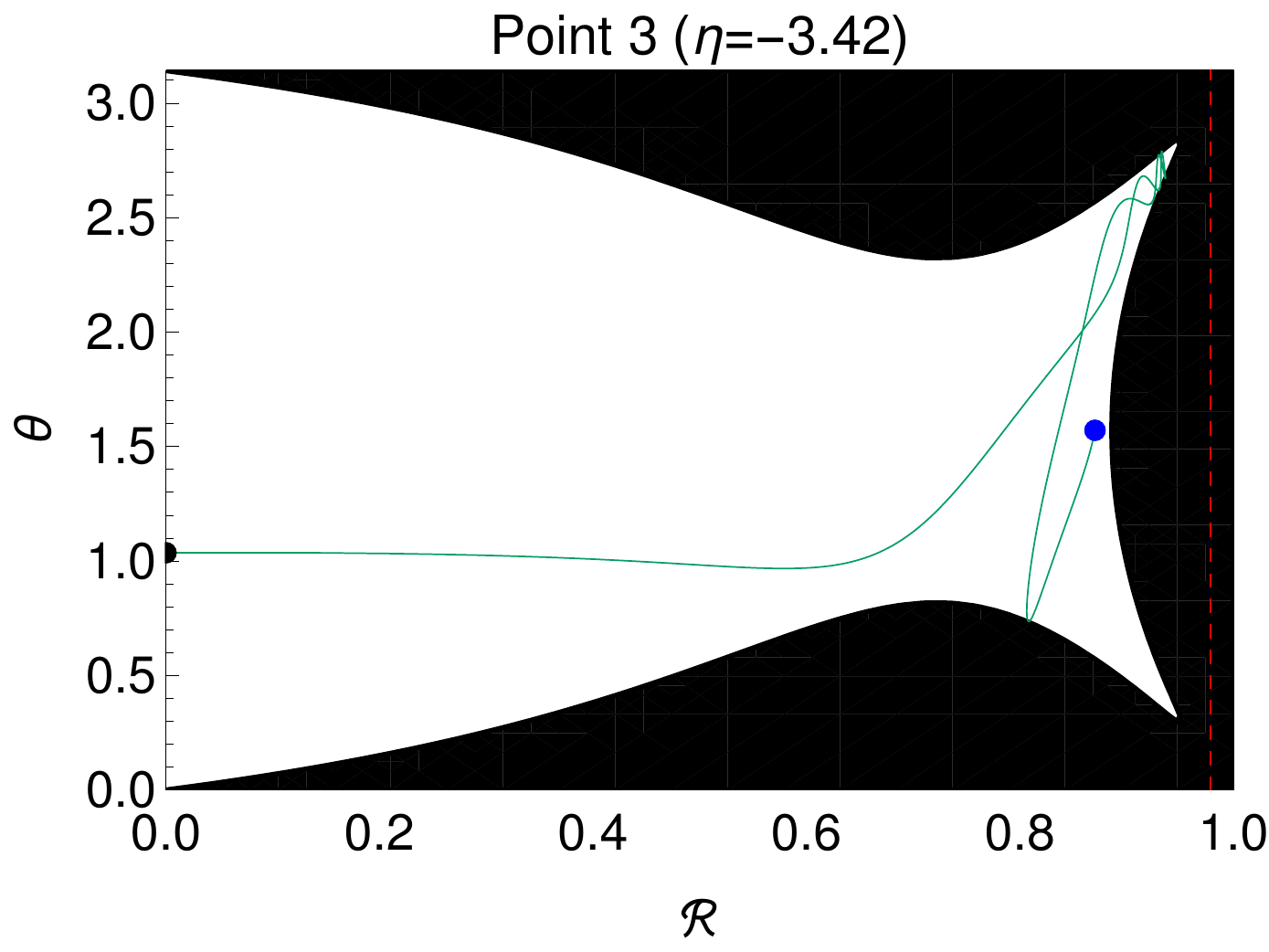}}
  \subfigure{\includegraphics[scale=0.42]{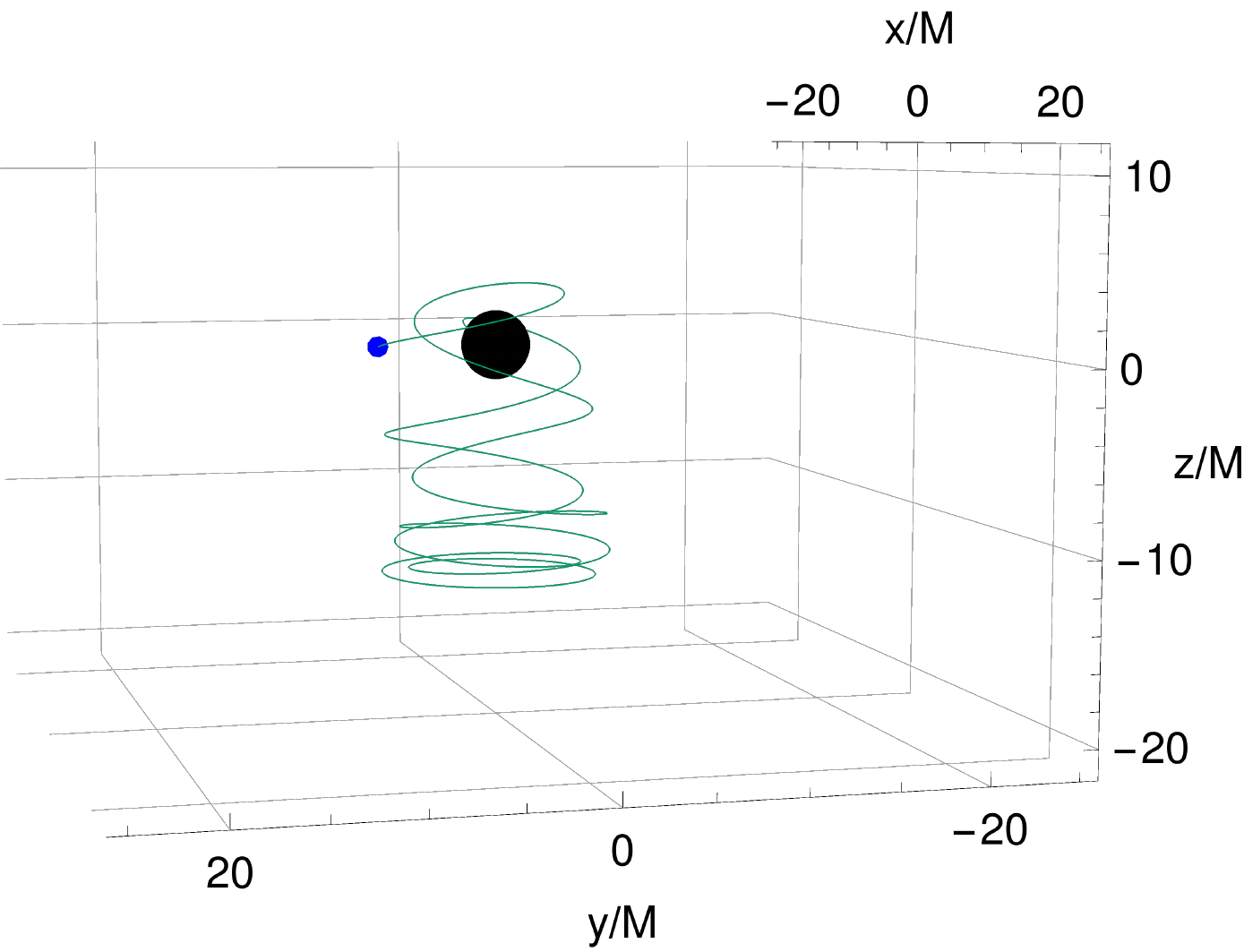}}\\
  \caption{Left: contour plots of the effective potential in the $\mathcal{R}-\theta$ plane, for the points selected in Fig.~\ref{ShadowB01_WP}. Right: trajectories described by the photon for the corresponding values of $\eta$ (displayed in the left panels).}
\label{Veff_off_equatorial}
\end{figure*}

The photon trajectory corresponding to point {\bf 1} shows how a north directed observation sees the southern hemisphere,  where the photon 
reaches the celestial sphere. 
 On the other hand, trajectory {\bf 2} shows  that  a slightly different observation direction is described by a photon that reaches the radial turning point (rather than the celestial sphere), whence it falls into the BH region. The same occurs for point {\bf 3} whose effective potential pushes the whole of the celestial sphere into the forbidden region, explaining why no colorful points can be found  in its vicinity. Besides the $r-\theta$ motion, Fig.~\ref{Veff_off_equatorial} also exhibits the nontrivial azimuthal motion, which adds to the  possible color mixing in the lensing  regions with chaotic patterns.

Before closing the analysis of Fig.~\ref{ShadowB01} there is one last point that merits discussion. The absence of equatorial LRs determines there is no shadow edge  on the equatorial plane, apart from the one  determined by the arbitrary  placement of the celestial sphere. How  about \textit{outside} the equatorial plane?  In asymptotically flat cases, the BH shadow edge in nonspherically symmetric spacetimes is determined, generically, by a set of bound photon orbits, dubbed \textit{fundamental photon orbits} (FPOs) in~\cite{Cunhaetal::2017}. For the Kerr case these are the familiar spherical orbits, which degenerate in the corotating and  counterrotating LRs on  the equatorial plane. Are there nonequatorial FPOs in the SMBH spacetime?

The answer to this question is  \textit{yes}. Since the SMBH does not admit separability, we have numerically computed such FPOs. These are illustrated in Fig.~\ref{FPO}, for the overcritical SMBH with $BM=0.2$. The corresponding points in the shadow edge, for each of the three illustrated  FPOs, are identified by the intersection between the vertical and horizontal dashed lines in Fig.~\ref{ShadowB01-middle} (with the same  color). It is worth remarking that the black FPO is planar; but it is polar rather than equatorial, and it is geometrically noncircular, as we  now  explain. First, observe that identical polar planar FPOs occur for any $\phi=\text{constant}$, due  to the axial symmetry and  always intersect the $z$ axis. Second, inspection of the $\phi=\text{constant}$ hypersurfaces of  the SMBH geometry~\eqref{line_el} shows that these sections are conformal to the Schwarzschild ones:
\begin{align}
\left. ds^2 \right|_{\phi=\text{cte.}}=\Lambda^2\left[-\left(1-\frac{2\,M}{r}\right)dt^2+\frac{dr^2}{\left(1-\frac{2\,M}{r}\right)}+r^2d\theta^2 \right].
\end{align}
Thus the usual Schwarzschild LR at $r=3M$ remains a geodesic here, but now the $\theta$ motion plays the role of the usual $\phi$ in  Schwarzschild and thus these are polar rather than equatorial orbits. These are our polar planar FPOs. Third,  since these polar orbits with $r=3M$ exist for every $\phi$, they span the $r=3M$, $t=constant$, 2-surface of the SMBH geometry~\eqref{line_el}:
\begin{align}
\left. ds^2 \right|_{t=\text{cte.},\ r=3M}=&(3M)^2\left[\Lambda^2(r,\theta)\,d\theta^2+
\frac{\,\sin^2\theta}{\Lambda^2(r,\theta)}\,d\phi^2\right].
\end{align}
This (topologically) 2-sphere is not round (but rather prolate, as the apparent horizon, Fig.~\ref{Horizon_embedding}). It has equatorial $\ell_{\rm eq}$ and polar $\ell_{\rm p}$ proper lengths:
\begin{align}
\frac{\ell_{\rm eq}}{2\pi(3M)} =& \frac{1}{1+\frac{(3MB)^2}{4}},
\\
\frac{\ell_{\rm p}}{2\pi(3M)} =  & 1+\frac{(3MB)^2}{8}.
\end{align}
Thus, we conclude that, in this sense, for $B\neq0$ the polar FPOs are not circular, but rather elongated in the magnetic field direction, since  $\ell_{\rm p}>\ell_{\rm eq}$.
Similar planar FPOs can be found in examples of di-BHs — see~\cite{Shipley::2016}.

The existence  of FPOs suggests that there are shadow edges that are independent of the celestial sphere location (unlike, say, the equatorial shadow edge). We also remark that FPOs also exists for the undercritical SMBHs and they determine the BH shadow edges shown in Fig.~\ref{Shadow1}.

\section{Final remarks}
\label{Final_remarks}
The SMBH spacetime first obtained by Ernst in 1976~\cite{Ernst::1976} as a solution of the Einstein-Maxwell (electrovacuum) equations, is interpreted as a Schwarzschild BH in a Melvin magnetic Universe. The latter is a sort of confining box, for light rays, that can only reach infinity if they have no angular momentum. The analysis of the null geodesics in this paper has shown that the SMBH provides a rich and novel case study for features of gravitational lensing and BH shadows. As a summary of our results we would like to emphasize the following features:
\begin{description}
\item[$\bullet$] The empty Melvin universe ($M=0\neq B$)~\cite{Melvin::1964} has an equatorial stable LR, which, moreover, due to a ``vertical" isometry, implies the existence of a \textit{tube of planar LRs.}
\item[$\bullet$] The SMBH family ($M\neq 0\neq B$) yields a novel example of BH spacetimes wherein the LRs total topological charge~\cite{Cunha::2017,Cunha::2020} is zero. This can be interpreted as the cancellation between the $w=-1$ Schwarzschild and  the $w=+1$ Melvin (asymptotics) topological charges.
\item[$\bullet$] The vanishing of the topological LR charge permits that, unlike for asymptotically flat~\cite{Cunha::2020} and asymptotically de Sitter or AdS~\cite{Wei::2020}  BHs, there can be asymptotically Melvin BHs without LRs.
\item[$\bullet$] The SMBHs family is then naturally divided into undercritical ($B<B_c$) and overcritical  ($B>B_c$) spacetimes, where $B_c$ is given by Eq.~\eqref{Bc}.  The latter describe BH spacetimes (regular on and outside a horizon) without LRs outside the horizon.
\item[$\bullet$] Performing ray-tracing, we have shown that the absence of  equatorial LRs yields  a \textit{panoramic shadow}: even a far away observer will see an (almost) $360\degree$  BH shadow, looking at any direction along the equatorial plane, as long  as the celestial sphere is sufficiently far away.
\item[$\bullet$] Despite the absence of LRs, there are nonequatorial (unstable) FPOs, that determine parts of the shadow edge, outside the equatorial plane. There are, in  particular planar, polar (but noncircular) photon orbits, which determine  the north and south pole of the shadow edge.
\item[$\bullet$]  Despite their simplicity,  both  the Melvin universe and the MSBH family yield chaotic lensing (and the latter also multiple shadows). Some insight into this behavior can be obtained by following illustrative orbits and analyzing the corresponding effective  potentials.
\item[$\bullet$] For small $BM$ (undercritical) SMBHs, the shadow becomes oblate, in a curious contrast with the intrinsic horizon geometry which becomes prolate. 
\end{description}

Although most of our results are unlikely to be astrophysically relevant (since Melvin asymptotics do not describe the real Universe), the shadow deformation for small magnetic fields could, potentially, describe those for an astrophysical BH immersed in a sufficiently strong poloidal magnetic field whose backreaction would have to be taken into account. Interestingly, the recent EHT results suggest that the M87* magnetic field is poloidal, rather than toroidal~\cite{EHT2021}.

Finally, our work suggests further studies. For instance, a deeper understanding of the chaotic patterns (and multiple shadows) as well as  their dependence on the celestial sphere would be desirable. These features are related to the existence of closed pockets and forbidden regions in the effective potential, leading to multiple turning points in the photons' trajectories. We have observed, for instance, that the time that a photon takes to reach the celestial sphere increases in the chaotic region due to the existence of several radial turning points, which has also been observed in other examples~\cite{KerrwSH_Shadow,KBHSH::2016,Shipley::2016,Bohn::2015}. As a second possible extension, a natural follow up with be to consider either the Reissner-Nordstr\"om-Melvin or the Kerr-Melvin solutions (see, $e.g.,$~\cite{Gibbons:2013yq}). Both of these acquire angular momentum, which will lead to frame-dragging effects in the lensing and shadows.

\acknowledgments
The authors thank Funda\c{c}\~ao Amaz\^onia de Amparo a Estudos e Pesquisas (FAPESPA),  Conselho Nacional de Desenvolvimento Cient\'ifico e Tecnol\'ogico (CNPq) and Coordena\c{c}\~ao de Aperfei\c{c}oamento de Pessoal de N\'{\i}vel Superior (Capes) - Finance Code 001, in Brazil, for partial financial support.
This work is supported by the Center for Research and Development in Mathematics and Applications (CIDMA) through the Portuguese Foundation for Science and Technology (FCT - Funda\c{c}\~ao para a Ci\^encia e a Tecnologia), references No. UIDB/04106/2020, No. UIDP/04106/2020, and No. BIPD/UI97/7484/2020. We acknowledge support  from the projects No. PTDC/FIS-OUT/28407/2017, No. CERN/FIS-PAR/0027/2019, and No. PTDC/FIS-AST/3041/2020. This work has further been supported by  the  European  Union's  Horizon  2020  research  and  innovation  (RISE) programme H2020-MSCA-RISE-2017 Grant No.~FunFiCO-777740. The authors would like to acknowledge networking support by the COST Action CA16104.

\end{document}